\documentclass[prd, aps, nofootinbib,preprintnumbers, showpacs,superscriptaddress,twocolumn]{revtex4-1} 
\usepackage{latexsym}
\usepackage{amssymb}
\usepackage{amsfonts}
\usepackage{amsmath}
\usepackage{bm}
\usepackage[dvips]{graphicx}
\usepackage{color}
\usepackage{times}
\usepackage{units}

\newcommand{\be}{\begin{equation}}
\newcommand{\ee}{\end{equation}}
\newcommand{\bea}{\begin{eqnarray}}
\newcommand{\eea}{\end{eqnarray}}
\newcommand{\beqn}{\begin{eqnarray*}}
\newcommand{\eeqn}{\end{eqnarray*}}
\newcommand{\ba}{\begin{align}}
\newcommand{\ea}{\end{align}}

\newcommand{\Msun}{\,M_{\odot}}

\newcommand{\cf}{{cf.}~}
\newcommand{\ie}{{i.e.,}~}
\newcommand{\eg}{{e.g.,}}

\newcommand{\km}{{\rm km}}

\def\lm{{\ell m}}

\def\h{\Delta}

\def\l{{\ell }}

\def\C{{\cal C}}

\def\k{\kappa}

\begin{document}


\title{Accurate numerical simulations of inspiralling binary neutron stars \\
       and their comparison with effective-one-body analytical models}


\author{Luca \surname{Baiotti}}
\affiliation{Institute of Laser Engineering, Osaka University, Suita, Japan}

\author{Thibault \surname{Damour}}
\affiliation{Institut des Hautes Etudes Scientifiques, Bures-sur-Yvette, France}
\affiliation{ICRANet, Pescara, Italy}

\author{Bruno \surname{Giacomazzo}}
\affiliation{Department of Astronomy, University of Maryland, College Park, MD USA}
\affiliation{Gravitational Astrophysics Laboratory, NASA Goddard Space Flight Center, Greenbelt, MD USA}

\author{Alessandro \surname{Nagar}}
\affiliation{Institut des Hautes Etudes Scientifiques, Bures-sur-Yvette, France}

\author{Luciano \surname{Rezzolla}}
\affiliation{Max-Planck-Institut f\"ur Gravitationsphysik, Albert-Einstein-Institut, Potsdam, Germany}
\affiliation{Department of Physics and Astronomy, Louisiana State University,Baton Rouge, LA, USA}

\begin{abstract}
Binary neutron-star systems represent one of the most promising
sources of gravitational waves. In order to be able to extract
important information, notably about the equation of state of matter
at nuclear density, it is necessary to have in hands an accurate
analytical model of the expected waveforms. Following our recent
work~\cite{Baiotti:2010}, we here analyze more in detail two
general-relativistic simulations spanning about $20$
gravitational-wave cycles of the inspiral of equal-mass binary neutron
stars with different compactnesses, and compare them with a tidal
extension of the effective-one-body (EOB) analytical model. The latter
tidally extended EOB model is analytically complete up to the 1.5
post-Newtonian level, and contains an analytically undetermined
parameter representing a higher-order amplification of tidal
effects. We find that, by calibrating this single parameter, the EOB
model can reproduce, within the numerical error, the two numerical
waveforms essentially up to the merger. By contrast, analytical models
(either EOB or Taylor-T4) that do not incorporate such a higher-order
amplification of tidal effects, build a dephasing with respect to the
numerical waveforms of several radians.
\end{abstract}


\maketitle

\section{Introduction}
\label{sec:intro}

Binary neutron-star inspirals are among the most promising and certain
target sources for the advanced versions of the currently operating
ground-based gravitational-wave (GW) detectors LIGO/Virgo/GEO. These
detectors will be maximally sensitive during the {inspiral} part of
the signal (around a GW frequency of $100$~Hz, \ie significantly below
the typical GW frequencies at merger, which are around $1000$~Hz). The
inspiral part of the signal will be influenced by tidal interaction
between the two neutron stars (NSs), which, in turn, encodes important
information about the equation of state (EOS) of matter at nuclear
densities. In other words, the detection of GWs emitted from
inspiralling NS in the LIGO/Virgo bandwidth could enable us to acquire
important information about the EOS of NS matter. However, besides 
getting sufficiently accurate GW data from advanced detectors, two
conditions must be fulfilled for the success of this program: (i)
obtaining a large enough sample of accurate {numerical simulations} of
inspiralling binary neutron stars (BNS); (ii) possessing a
sufficiently accurate {\it analytical model} of inspiralling BNS,
allowing the extrapolation of the finite set of numerical simulations
to the multi-parameter space of possible GW templates. Extending the
work recently reported in~\cite{Baiotti:2010}, we here address issues
and provide useful progress on both of them. In essence, we will
present the results of general-relativistic simulations spanning about
$20$ gravitational-wave cycles of the inspiral of equal-mass BNSs, and
show how a suitably calibrated effective-one-body (EOB) analytical
model of tidally interacting BNS systems enables us to accurately
reproduce the numerically simulated inspiral waveform.

Numerical simulations of merging BNSs in full general relativity have
a long history (see the Introduction of~\cite{Baiotti08} for a brief
review), and the first merger to a hypermassive neutron star (HMNS) was
computed more than ten years ago~\cite{Shibata99d}. However, it is
only in recent years and with the use of more advanced and accurate
numerical algorithms that it has been possible to obtain a more
precise and robust description of this process and to include
additional physical ingredients such as magnetic fields and realistic
EOSs. In particular the use of adaptive mesh refinement
techniques~\cite{Anderson2007,Baiotti08,Yamamoto2008} made it possible
to use very high resolutions, increasing not only the level of
accuracy, but giving the possibility, for example, to compute the full
evolution of the HMNS up to black-hole (BH) formation~\cite{Baiotti08} or
to investigate in detail the development of hydrodynamical
instabilities at the time of the merger~\cite{Baiotti08}. Also the
numerical convergence properties of BNS simulations have been
studied only recently~\cite{Baiotti:2009gk}, providing for the
first time evidence of the level of accuracy that it is now possible
to achieve in the generation of GW templates from these
sources. Several groups are now able to simulate BNSs using more
realistic EOSs (see, \eg~\cite{Read:2009b,Kiuchi2009,Kiuchi2010} and
references therein) and to assess the possibility to measure their
effects in the GW signals. In the last two years three different
groups were also able to perform for the first time the simulations of
magnetized BNSs~\cite{Anderson2008,Liu:2008xy,Giacomazzo:2009mp}. One
conclusion already reached is that no effect of the magnetic field can
be measured in the inspiral waveforms~\cite{Giacomazzo:2009mp}, while
the role of the magnetic field in the post-merger phase has been
recently investigated in~\cite{Giacomazzo:2010} as well as its role
in the emission of relativistic jets after the collapse to 
BH~\cite{Rezzolla:2011}. Because of their possible connection with
the production of short gamma-ray bursts (GRBs), numerical simulations
have also investigated in detail the formation of massive tori and
their dependence on the initial mass and mass ratio of the binary (see 
\eg~\cite{Rezzolla:2010}) as well as on the EOS used
(see~\cite{Kiuchi2009,Kiuchi2010} and references therein).

On the other hand, the program of developing an analytical description
within general relativity of tidally-interacting binary systems has
been initiated only recently~\cite{Flanagan08, Hinderer08,
  Damour:2009, Binnington:2009bb, Hinderer09, Damour:2009wj,
  Pannarale2011}. Overall, this work has brought to light two
surprising results. First, that the dimensionless expression $k_\ell$
(Love number) in the (gravito-electric) tidal polarizability parameter
$G\mu_\ell \equiv 2 k_\ell R^{2\ell +1}/(2\ell -1)!!$ measuring the
relativistic coupling (of multipolar order $\ell$) between a NS of
radius $R$ and the external gravitational field in which it is
embedded strongly decreases with the compactness parameter ${\cal
  C}\equiv GM/(c^2 R)$ of the NS~\cite{Damour:2009,
  Binnington:2009bb}\footnote{As a consequence, for a given EOS, the
  Love numbers of a typical (${\cal C}\simeq 0.15$) NS are found to be
  about $4$ time smaller than their corresponding Newtonian estimates,
  that assume ${\cal C}\to 0$.}. Second, a recent comparison between
a numerical computation of the binding energy of quasi-equilibrium
circular sequences of BNS systems~\cite{Uryu:2009ye} and the EOB
description of tidal effects~\cite{Damour:2009wj} suggests that
high-order (beyond the first order) post-Newtonian (PN) corrections to
tidal effects tend to {significantly increase} (typically by a factor
of order two) the effective tidal polarizability of NSs.

The main aim of this paper is to present a detailed comparison between
waveforms computed from the tidal-completed EOB analytical model of
Ref.~\cite{Damour:2009wj} and waveforms from BNS simulations comprising between
$\simeq 20$ and $22$ GW cycles of inspiral~\cite{Baiotti:2010}. More specifically, 
we will follow Ref.~\cite{Damour:2009wj}, which has proposed a new way of
analytically describing the dynamics of tidally interacting BNSs,
whose validity is not a priori limited (like the purely PN-based
descriptions used in, \eg~\cite{Flanagan08}) to the low-frequency part
of the GW signal, but may be extended to higher frequencies,
essentially up to the merger. The proposal of
Ref.~\cite{Damour:2009wj} consists in extending the EOB
method~\cite{Buonanno:1998gg,Buonanno00a,Damour:2001tu}, which has
recently shown its ability to accurately describe the GW waveforms
emitted by inspiralling, merging, and ringing binary black holes
(BBHs)~\cite{Damour:2009kr,Buonanno:2009qa}, by incorporating tidal
effects in it. We will improve the tidally-extended EOB model of
Ref.~\cite{Damour:2009wj} (which already contained the 1PN
contributions to the dynamics) by incorporating the 1PN contributions
to the waveform (from \cite{Vines2011}), as well as the waveform tail
effects (from~\cite{Blanchet:1992br,Damour:2007yf}).

The paper is organized as follows. In Sec.~\ref{sec:numerics} we
present in detail our numerical simulations, briefly reviewing our
numerical setup, discussing the dynamics of the binaries, and
presenting the main features of the
waveforms. Section~\ref{sec:analytics} deals instead with the
analytical models of the binary dynamics and of waveforms that include
tidal interaction (either PN-based or EOB-based). 
Sec.~\ref{sec:Qomega} introduces some tools, notably a certain
intrinsic representation of the time evolution of the GW frequency,
which is useful for doing the
numerical-relativity/analytical-relativity (NR/AR) comparison.
Section~\ref{sec:errors} discusses the various errors that affect the
NR phasing. The NR/AR comparison is carried out in
Sec.~\ref{sec:results}. We finally present a summary of our findings
in Sec.~\ref{sec:end}. Two appendices give additional technical
details on the use of the waveforms from the numerical-relativity
simulations.

We use a spacelike signature $(-,+,+,+)$ and (unless explicitly said
otherwise) a system of units in which $c=G=M_\odot=1$. Greek indices
are taken to run from $0$ to $3$, Latin indices from $1$ to $3$.

\section{Numerical-relativity simulations}
\label{sec:numerics}

\subsection{Numerical setup}
\label{numsetup}

The numerical simulations were performed with the set of codes {\tt
  Cactus}-{\tt Carpet}-{\tt Whisky}~\cite{Goodale02a,
  Schnetter-etal-03b, Pollney:2007ss, Baiotti03a, Baiotti04}. The
reader is referred to these references for the description of the
details of the implementations and of the tests of the codes. Since in
this work we use the same gauges and numerical methods already applied
and explained in~\cite{Baiotti08,Baiotti:2009gk}, we also refer the
reader to these articles for more detailed explanations of the setup
only briefly recalled below.

\begin{table*}[ht]
  \caption{\label{tab:models}Properties of the binary NS initial
    data. From left to right the columns show: the name of the model,
    the total baryonic mass \ensuremath{M^{\rm bar}_{{\mathrm{tot}}}}
    of the system, the total (initial) Arnowitt-Deser-Misner (ADM)
    mass \ensuremath{M_{_{\mathrm{ADM}}}} of the system, the total
    (initial) angular momentum \ensuremath{J}, the initial orbital
    frequency \ensuremath{f_\mathrm{orb}}, the initial maximum
    rest-mass density \ensuremath{\rho_\mathrm{max}}, the mean radius
    \ensuremath{\bar r} of each star, the axis ratio \ensuremath{\bar
      A} of each star, the individual ADM mass $M^{\infty}$ of each
    star as considered in isolation at infinity, the compactness
    ${\cal C}^{\infty}=M^\infty_{NS}/R^\infty_{NS}$ of each star as
    considered in isolation at infinity, the corresponding
    (quadrupolar) dimensionless Love number $k_2$ and tidal constant
    $\k_2^T$ as defined in Ref.~\cite{Damour:2009wj} (see also
    Eq.~\eqref{eq:def_kT} below). The mean radius is defined as
    \ensuremath{\bar r \equiv (r_\vdash + r_\dashv + r_\perp +
      r_\mathrm{pol}) / 4}, where \ensuremath{r_\vdash} and
    \ensuremath{r_\dashv} are the (coordinate) radii of the star
    parallel to the line connecting the stars, \ensuremath{r_\perp} is
    the radius in the equatorial plane perpendicular to that line, and
    \ensuremath{r_\mathrm{pol}} is the radius perpendicular to the
    equatorial plane. The axis ratio is defined as the ratio between
    the mean radius parallel to the line connecting the stars and the
    mean radius in the plane perpendicular to that line, namely
    \ensuremath{\bar A \equiv (r_\perp + r_\mathrm{pol}) / (r_\vdash
      + r_\dashv)}. The values of ${f_\mathrm{orb}}$, ${\bar r}$,
    ${\bar A}$, $M^{\infty}$, and ${\cal C}^{\infty}$ are computed
    with the \texttt{LORENE} code, the values of ${M^{\rm
        bar}_{{\mathrm{tot}}}}$, ${M_{_{\rm ADM}}}$, $J$, and
    $\rho_\mathrm{max}$ are instead measured on the Cartesian grid by
    the \texttt{Whisky} code, and those of $k_2$ (and $\kappa_2^T$)
    are computed according to Ref.~\cite{Damour:2009}.}
\begin{center}
\begin{ruledtabular}
\footnotesize{\begin{tabular}[t]{lccccccccccc}
Model & ${M^{\rm bar}_{{\mathrm{tot}}}}$ & ${M_{_{\rm ADM}}}$
    & {$J/10^{49}$} & ${f_\mathrm{orb}}$ &
    {${\rho_\mathrm{max}}/10^{14}$} & ${\bar r}$ & ${\bar
      A}$ & $M^{\infty}$ &
    ${\cal C}^{\infty}$ & $k_2$ & $\kappa_2^T$ \\ 
    & {(${\Msun}$)} & {(${\Msun}$)} &  {(${\unit{g\,cm^2/s}}$}) & {(${\unit{Hz}}$)} &  {(${\rm g/cm}^3$)} & {(${\rm km}$)} & & {(${\Msun}$)} & \\
    \hline\hline
\texttt{M2.9C.12}   & $2.8899$ & $2.6925 $ &  $7.1747$ & $188.52$ & $4.60$ & $14.2$ & $0.97$ & $1.359$ &$0.1196$ & $0.09719$ & $496.09$ \\    
\texttt{M3.2C.14}   & $3.2504$ & $2.9966 $ &  $8.5558$ & $197.03$ & $5.93$ & $13.2$ & $0.97$ & $1.514$ &$0.1399$ & $0.07894$ & $183.81$
\\    
\end{tabular}
}
\end{ruledtabular}
\end{center}
\end{table*}

In essence, we evolve a conformal-traceless ``$3+1$'' formulation of
the Einstein equations in which the spacetime is decomposed into
three-dimensional spacelike slices, described by a metric
$\gamma_{ij}$, its embedding in the full spacetime, specified by the
extrinsic curvature $K_{ij}$, and the gauge functions $\alpha$ (lapse)
and $\beta^i$ (shift), which specify a coordinate frame (see
Ref.~\cite{Pollney:2007ss} for details on the latest implementation of
the Einstein equations in the code). For the evolution of the matter,
the {\tt Whisky} code implements the flux-conservative formulation of
the general-relativistic hydrodynamics equations proposed by the
Valencia group~\cite{Banyuls97}. Among its important features is that the
set of conservation equations for the stress-energy tensor
$T^{\mu\nu}$ and for the matter current density $J^\mu$ are written in
hyperbolic, first-order, and flux-conservative form (see
Ref.~\cite{Baiotti08} for details on the latest implementation of the
hydrodynamics equations in the code).

As initial data we use quasi-equilibrium binaries generated with the
multi-domain spectral-method code \texttt{LORENE} developed at the
Observatoire de Paris-Meudon~\cite{Gourgoulhon01}. For more
information on the code and its methods, the reader is referred to the
\texttt{LORENE} web pages \cite{Lorene}. In particular, we use
irrotational configurations, defined as having vanishing vorticity
and obtained under the additional assumption of a conformally flat
spacetime metric~\cite{Gourgoulhon01}. The EOS assumed for the initial
data is in all cases the polytropic EOS 
\begin{equation}
\label{eq:poly}
p = K\,\rho^\Gamma\,,
\end{equation}
where $p$ and $\rho$ are the pressure and the rest-mass
(baryonic-mass) density, respectively. The chosen adiabatic index is
${\Gamma = 2}$, while the polytropic constant is $K \simeq 123.6$ (in
units where $c=G=\Msun=1$).
For this particular EOS, the allowed maximum baryonic mass for an
individual stable NS is ${\unit{2.00}{\Msun}}$, thus leading to a
maximum compactness $M_{_{\rm ADM}}/R \simeq 0.25$. The initial
coordinate separation of the stellar centers in all cases is $d =
\unit{60~km}$.

The physical properties of the two binaries considered here are
summarized in Table~\ref{tab:models}, where we have adopted the
following naming convention: \texttt{M\%C\#}, with \texttt{\%} being
replaced by the rounded total baryonic mass ${M^{\rm
    bar}_{\mathrm{tot}}}$ of the binary NS system and \texttt{\#} by
the compactness. As an example, \texttt{M2.9C.12} is the binary with
total baryonic mass $M^{\rm bar}_{\mathrm{tot}}=2.8899\,\Msun$ and
compactness ${\cal C}=0.1196$. We note that at least as far as the
tidal effects are concerned, the most important difference in the two
sets of initial data is represented by the compactness, which is
smaller in the binary \texttt{M2.9C.12} than in the binary
\texttt{M3.2C.14}. Note that the dimensionless EOB parameter
$\kappa_2^T$ measuring the strength of the (conservative) quadrupolar
interaction is nearly three times larger for $\C =0.12$, than for
$\C=0.14$.

The initial data is then evolved either using the (isentropic) polytropic
EOS~\eqref{eq:poly} or using the (non-isentropic)  ``ideal-fluid'' EOS defined by
the condition
\begin{equation}
\label{eq:IF}
p = \rho\, \epsilon(\Gamma-1),
\end{equation}
where $\epsilon$ is the specific internal energy and
$e=\rho(1+\epsilon)$ is the total energy density. Although these EOSs
are idealized, they provide a reasonable approximation of the dynamics
of NSs during the inspiral, so that we expect that the use of
realistic EOSs (with similar compactnesses) would not change the main
qualitative conclusions of this work. A detailed discussion of the
consequences of using either EOS will be presented in
Sec.~\ref{sec:errors}.

As mentioned above, the use of adaptive mesh-refinement techniques
allows us to reach a considerable level of precision and for this we
use the \texttt{Carpet} code~\cite{Schnetter-etal-03b} that implements
a vertex-centered adaptive-mesh-refinement scheme adopting nested
grids with a $2:1$ refinement factor for successive grid levels. We
center the highest resolution level around the peak in the rest-mass
density of each star. This represents our rather basic form of
adaptive-mesh refinement. The timestep on each grid is set by the
Courant condition (expressed in terms of the speed of light) and so by
the spatial grid resolution for that level; the typical Courant
coefficient is set to be $0.35$. The time evolution is carried out
using fourth-order accurate Runge-Kutta integration steps. Boundary
data for finer grids are calculated with spatial prolongation
operators employing fifth-order polynomials and with prolongation in
time employing second-order polynomials.

In the results presented below we have used $6$ levels of mesh
refinement with the finest grid resolution of $\h_{\rm
  min}=0.12\,M_{\odot}=0.177\,\km$ and the coarsest (or wave-zone)
grid resolution of $\h_{\rm max}=3.84\,M_{\odot}=5.67\,\km$. Each star
is completely covered by the finest grid, so that the high-density
regions of the stars are tracked with the highest resolution
available. The refined grids are then moved by tracking the position
of the maximum of the rest-mass density as the stars orbit, and are
finally merged when they overlap. In addition, a set of refined but
fixed grids is set up at the center of the computational domain so as
to capture the details of the Kelvin-Helmholtz instability
(\cf~\cite{Baiotti08}). The finest of these grids extends to
$r=7.5\,M_{\odot}=11\,\km=5.52 M$ for model {\tt M2.9C.12} and
$=4.95M$ for model {\tt M3.2C.14} (here and in the following $M$
denotes the gravitational mass of the system at infinite separation,
namely the sum of the gravitational masses of each NS as computed
individually in isolation, \ie $M \equiv 2M_{NS}^\infty$ in the
notation of Table~\ref{tab:models}). A single grid-resolution covers
then the region between $r=150\,M_{\odot}=221.5\,\km$ and
$r=514.56\,M_{\odot}=755.24\,\km$ (or $r=378.63M$ for {\tt M2.9C.12}
and $r=339.87M$ for {\tt M3.2C.14}), in which our wave extraction is
carried out. The resolution is here $\h=3.84\,M_{\odot}=5.67\,\km$ and
thus more than sufficient to accurately resolve the gravitational
waveforms that have initially a wavelength of about $720\,\km$.

A reflection symmetry condition across the $z=0$ plane and a
$\pi$-symmetry condition\footnote{Stated differently, we evolve only
the region $\{x\geq 0,\,z\geq 0\}$ applying a 180-degree
rotational-symmetry boundary condition across the plane at $x=0$.}
across the $x=0$ plane are used. A number of tests have been performed
to ensure that both the hierarchy of the refinement levels described
above and the resolutions used yield results that are numerically
consistent although not always in a convergent regime at the time of
merger (see the detailed discussion in Ref.~\cite{Baiotti:2009gk}).

\subsection{Overall matter dynamics and gravitational waveforms}

\begin{figure*}[t]
\begin{center}
\includegraphics[width=0.45\textwidth]{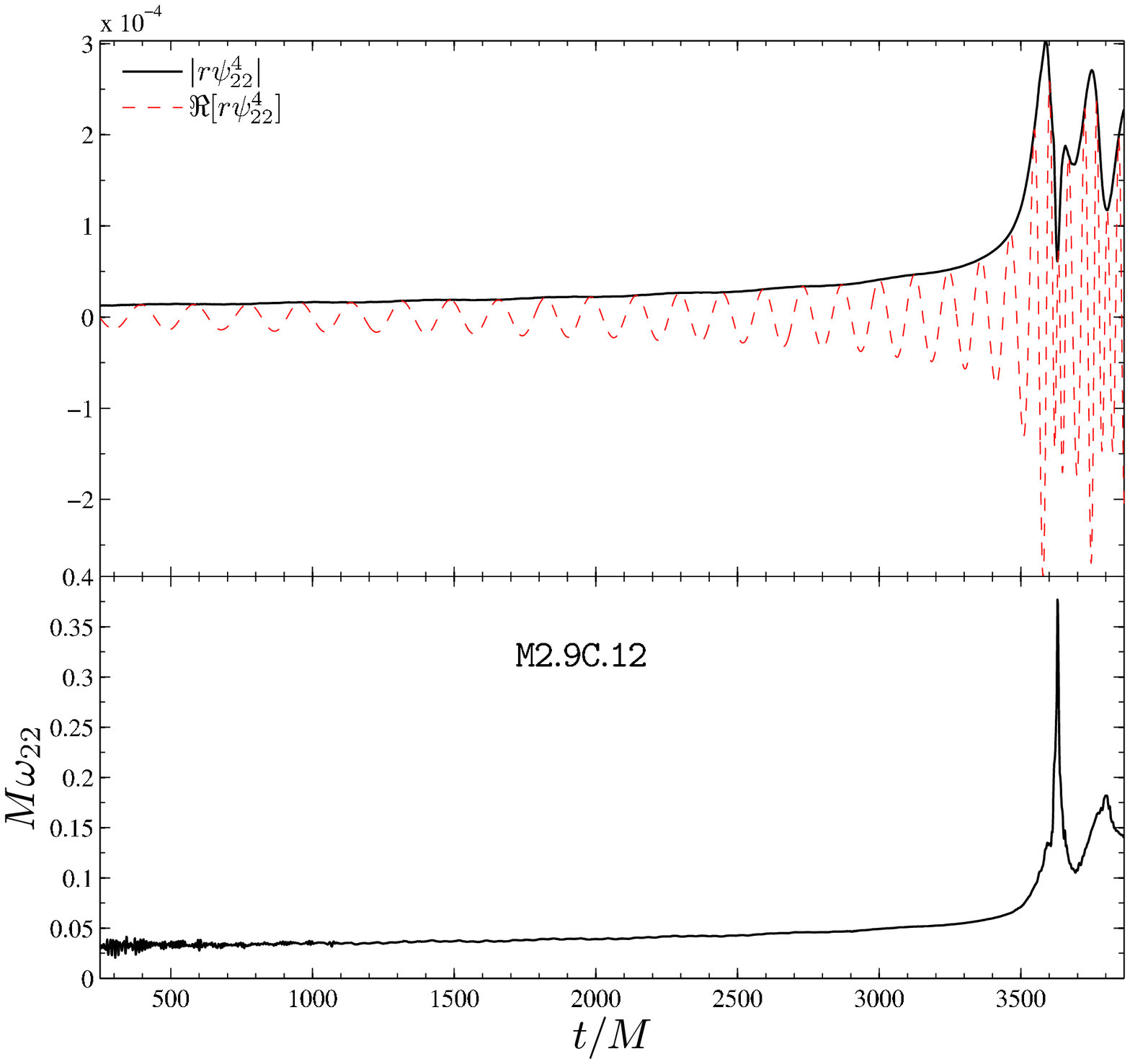}\hspace{10 mm}
\includegraphics[width=0.45\textwidth]{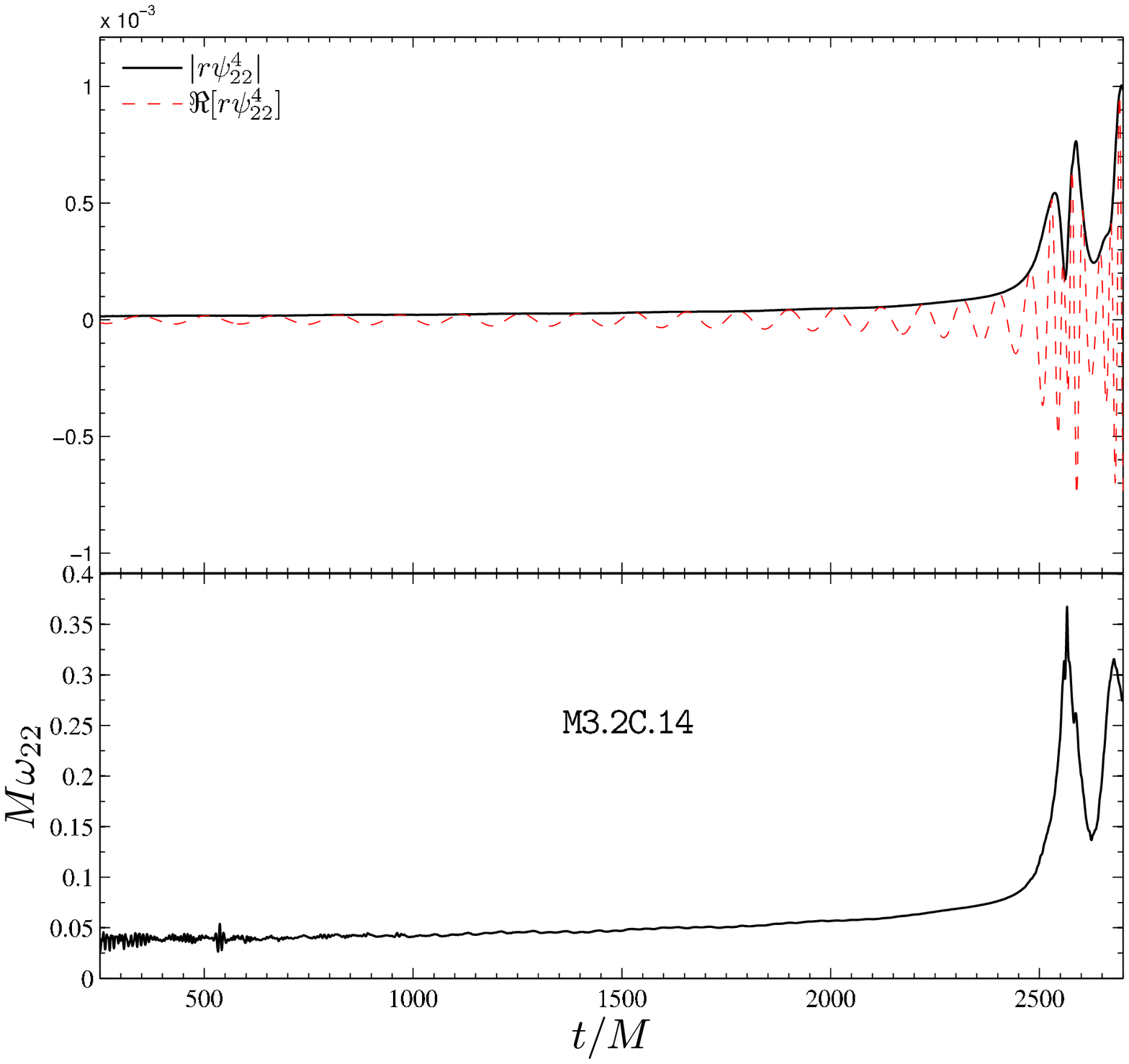}
\caption{\label{fig:fig_Psi4_waves}Curvature gravitational waveforms 
  $r\psi_4^{22}$ (upper panels) and their instantaneous frequency 
  $M\omega_{22}$ (lower panels) for the {\tt M2.9C.12}
  (left) and {\tt M3.2C.14} (right) models. In both cases, the
  observer's (coordinate) extraction radius is $r_{\rm obs}=500\Msun$;
  this corresponds to $r_{\rm obs }/M= 184.3$ for {\tt M2.9C.12} and
  $r_{\rm obs}/M=165.1$ for {\tt M3.2C.14}.}
  \end{center}
\end{figure*}

We next briefly recall the physical properties of BNS inspiral and
merger as discussed in Refs.~\cite{Baiotti08,Baiotti:2009gk}. The
inspiral proceeds at higher and higher frequencies until the time of
the merger, just before which the stars decompress because of the
tidal force. At the time of the merger, a Kelvin-Helmholtz instability
develops in the shearing layer formed by the colliding stars, which
could lead to an exponential growth of magnetic fields if these are
present~\cite{Price06, Obergaulinger10}; such a large growth was not
found in recent related
works~\cite{Giacomazzo:2009mp,Giacomazzo:2010}, and no magnetic fields
are included in the simulations reported here. If the total mass of
the system is sufficiently large, the merged object immediately
collapses to a Kerr BH, while, for smaller masses the merger remnant
is a HMNS in a metastable equilibrium. Because of the excess angular
momentum, the HMNS is also subject to a bar deformation, being
responsible for a copious emission of gravitational radiation with
peak amplitudes that are comparable or even larger than those at the
merger (\cf Ref.~\cite{Baiotti08}). As the bar-deformed HMNS loses
energy and angular momentum via GWs, it contracts and spins up, thus
further increasing the losses. The process terminates when the
threshold to the collapse to BH is crossed and the HMNS then rapidly
produces a rotating BH surrounded by a torus of hot and high-density
material. Although this post-merger evolution of the binary is of
great interest and is likely to yield a wealth of physical
information, it will not be further considered in the present work,
which is instead focussed on the analytical modelling of the inspiral
phase, up to the merger.

The GW signal is extracted at different surfaces of constant
coordinate radius $r_{\rm obs}$ by means of two distinct methods. The
first one is based on the measurements of the non-spherical
gauge-invariant perturbations of a Schwarzschild
BH~\cite{Nagar05,Nagar06}. The second and independent one uses instead
the Newman-Penrose formalism so that the GW (metric) polarization
amplitudes $h_+$ and $h_\times$ are then related to $\psi_4$ by (see
Sec.~IV of Ref.~\cite{Baiotti08} for details of the Newman-Penrose
scalar extraction in our setup)
\begin{equation}
\ddot{h}_+ - {\rm i} \ddot{h}_\times =
\psi_4 = \sum_{\ell=2}^{\infty}\sum_{m=-\ell}^{\ell} 
\psi_4^{\ell m}\;_{-2}Y_{\ell m}(\theta,\phi),
\end{equation} 
where we have introduced the (multipolar) expansion of $\psi_4$ in
spin-weighted spherical harmonics~\cite{Goldberg:1967} of spin-weight
$s=-2$. The coordinate extraction radius is $r_{\rm obs}=500\Msun$ for
both models, which corresponds to $r_{\rm obs }/M= 184.3$ for {\tt
  M2.9C.12} and to $r_{\rm obs}/M=165.1$ for {\tt M3.2C.14}. The top
panels of Fig.~\ref{fig:fig_Psi4_waves} summarizes most of the
information related to the $\ell=m=2$ curvature waveforms
$\psi_4^{22}$ for the {\tt M2.9C.12} model (left panels) and for the
{\tt M3.2C.14} model (right panels). The top panels of the figures
show together the modulus and the real part of the waveform; the
bottom ones, illustrate the behavior of the instantaneous GW
(curvature) frequency $M\omega_{22}$. Note that the inspiral waveform
of {\tt M2.9C.12} contains about 22 GW cycles, while that of {\tt
  M3.2C.14} contains about $20$ GW cycles. To fix conventions, let us
recall that we write the waveform as a complex number according to
\begin{equation}
\label{eq:def_psi4} 
\psi_4^{\lm}=|\psi_4^{\lm}| e^{-{\rm i}\phi_{\lm}}\ ,
\end{equation} 
so that the instantaneous (curvature) GW frequency is simply defined
as $\omega_{\lm}\equiv\dot{\phi}_\lm$. After the initial junk
radiation (\cf Ref.~\cite{Baiotti:2008nf}) that is responsible for a
spike in the modulus around $t=200M$ together with incoherent
oscillations in the frequency, the complex $\psi_4^{22}$ waveform
becomes circularly polarized (as expected for circularized inspiral),
with a modulus that grows monotonically in time up to the merger (see
upper panels of Fig.~\ref{fig:fig_Psi4_waves}).

\begin{figure*}[t]
\begin{center}
\includegraphics[width=0.45\textwidth]{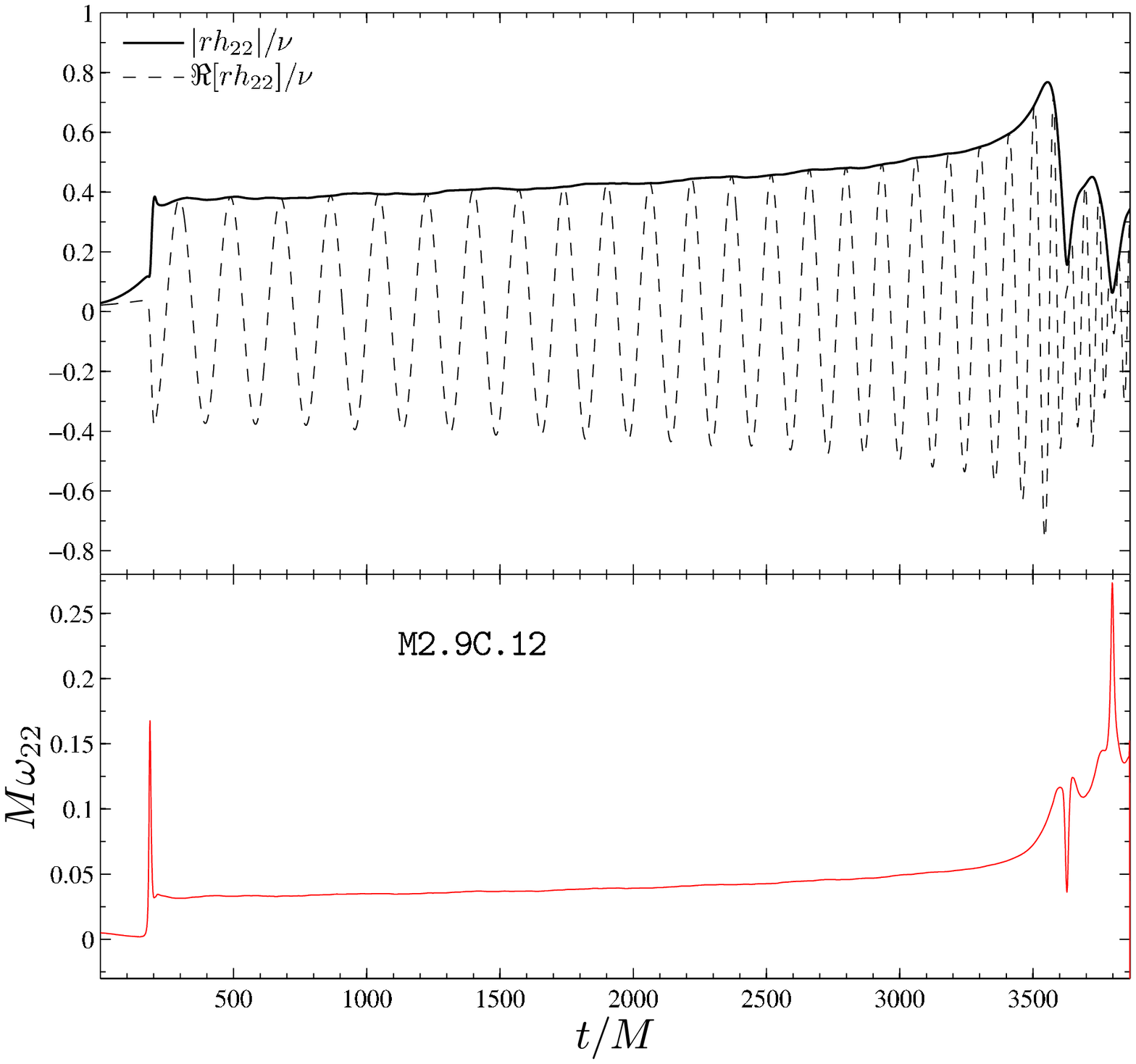}\hspace{10 mm}
\includegraphics[width=0.45\textwidth]{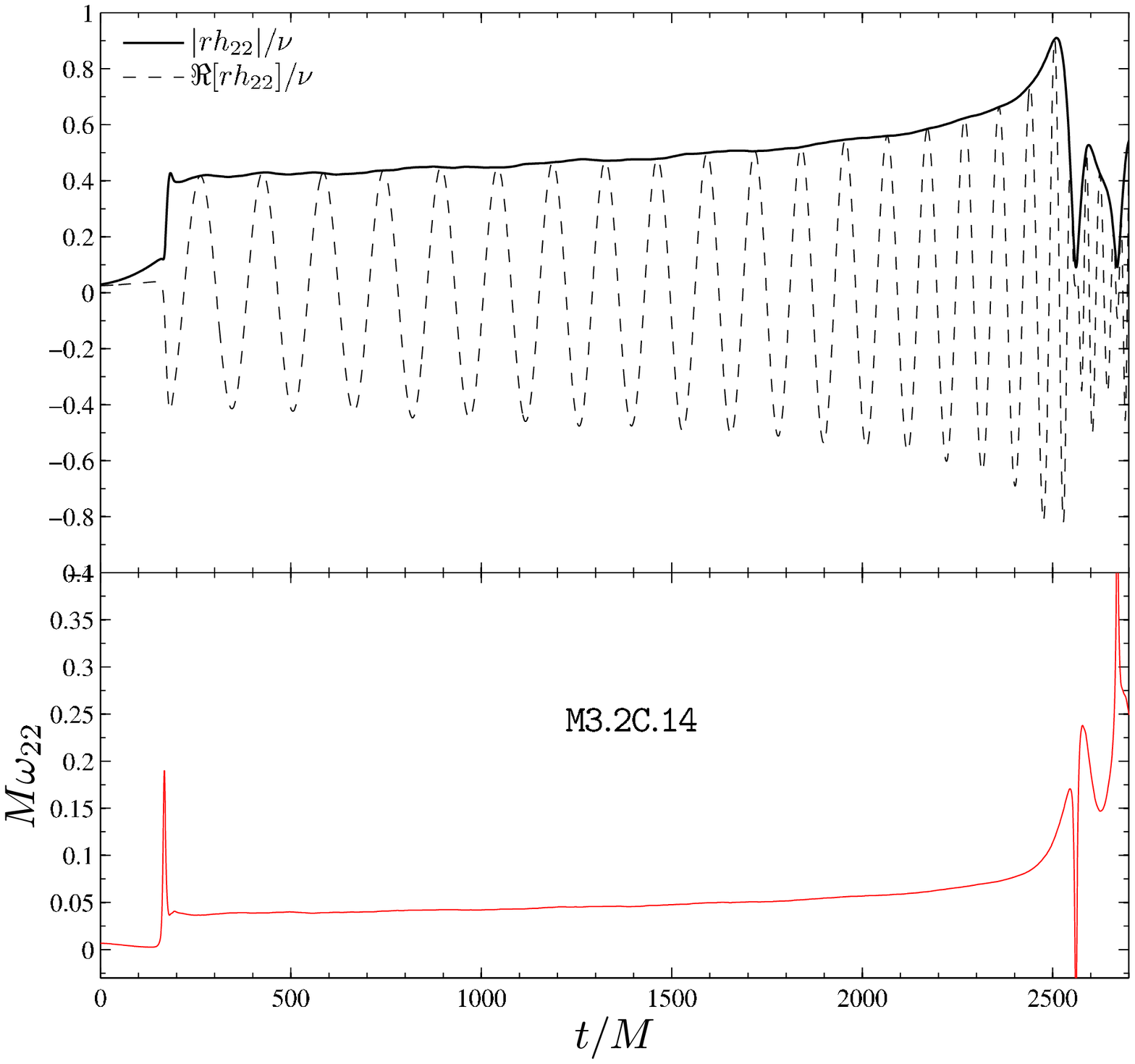}
\caption{\label{fig:fig_metric_waves} Metric gravitational
  waveforms $rh_{22}$ and frequencies (upper panels) and the corresponding istantaneous
 frequency $M\omega_{22}$ (lower panels) obtained from the (double) time-integration 
of the curvature waveforms of Fig.~\ref{fig:fig_Psi4_waves} [see Eq.~\eqref{eq:int_h_infty}].
The left panels refer to model {\tt M2.9C.12}, the right panels to model 
{\tt M3.2C.14}. The fact that the waveform modulus grows monotonically
without evident spurious oscillations is the indication of the reliability 
of the determination of the integration constants. See text for details.}
  \end{center}
\end{figure*}

The matter dynamics is reflected in the behavior of the frequency: for
both models we clearly see that $\omega_{22}$ grows monotonically
during the inspiral phase, until it reaches a maximum around the
``merger". In this work, we phenomenologically define the ``NR
merger'' as the instant when the modulus of the metric waveform
$h_{22}$ (see below) reaches its (first) maximum. Roughly speaking, in
our simulations the ``dynamical range'' of the dimensionless GW
frequency parameter $M\omega_{22}$ during the inspiral (\ie before the
merger) is $0.015\lesssim M\omega_{22}\lesssim 0.15$. Note that, if we
were considering a conventional $1.4\Msun-1.4\Msun$ BNS system, we
would then have the correspondence $f_{\rm GW}/100{\rm Hz}\simeq 115.4
M\omega_{22}$ so that $M\omega_{22}=0.015$ corresponds to $f_{\rm
  GW}\approx 173.1$~Hz, while $M\omega_{22}=0.15$ corresponds to
$f_{\rm GW}\approx 1731$~Hz.

In order to perform direct comparisons with (resummed) analytical
waveforms and since the resummations used in the EOB method have been
developed (and tested) mainly for metric waveforms, we derived
the metric waveform by a (double) time-integration of 
the $\psi_4^{22}$ waveform (the so-obtained metric waveform was 
found to be more accurate than the output of the gauge-invariant 
perturbation scheme). We recall that the metric waveform is also 
expanded in spin-weighted spherical harmonics with the 
following convention
\begin{equation}
h_+ - {\rm i} h_\times = \sum_{\ell=2}^{\infty}\sum_{m=-\ell}^{\ell}
h_{\ell m}\;_{-2}Y_{\ell m}(\theta,\phi),
\end{equation}
so that the metric multipoles $h_{\lm}$ at time $t$ can be obtained from 
 $\psi_{4}^{\lm}$ by double time-integration as
\begin{equation}
\label{eq:int_h_infty}
h_{\ell m}(t) = \int_{-\infty}^{t} dt' \int_{-\infty}^{t'} dt'' \psi_4^{\ell m}(t'').
\end{equation}
This expression assumes that one knows the curvature waveform on the
infinite time interval $(-\infty,t]$. Since, however, the simulated
curvature waveform does not start at an infinite time in the past,
but at a finite (conventional) time $t=0$, one has to find a way of
determining {two} (complex) {integration constants} accounting from
the GW emission from infinite time to our present starting time. To
do so, we use here an improved version of the fit procedure of
Ref.~\cite{Damour:2008te}, which is presented in detail in
Appendix~\ref{appendix_A}. Figure 2 shows the result of this
process, with the left panels referring to model {\tt M2.9C.12} and
the right ones to model {\tt M3.2C.14}. Note that the waveforms
displayed in these figures are obtained from simulations with: (i)
the non-isentropic (ideal fluid) EOS; (ii) the highest available
resolution; and (iii) an extraction radius of $500\Msun$. These will
be taken as our fiducial ``target'' waveforms for our NR/AR
comparisons, and we will refer to them in the following with the
label ${\rm IF_{HR}}500$. The numerical uncertainty on these target
waveforms will be estimated in Sec.~\ref{sec:errors} below.

\section{Analytical models}
\label{sec:analytics}

We recall below some basic information relative to the EOB-based and
PN-based descriptions of the binary dynamics and waveforms that
include tidal effects. We follow here the general discussion of
Ref.~\cite{Damour:2009wj}, to which we refer the reader for more
details. We consider successively: (i) the resummed EOB description of 
the conservative dynamics, (ii) the resummed  EOB description of the waveform, 
and (iii) one of the non-resummed (\ie PN expanded) 
descriptions of the phasing.

\subsection{Effective-one-body description of the conservative dynamics}
\label{sbsc:eob}

The EOB formalism~\cite{Buonanno:1998gg,Buonanno00a,Damour:2001tu}
replaces the PN-expanded two-body interaction Lagrangian (or
Hamiltonian) by a resummed Hamiltonian, of a specific form, which
depends only on the relative position and momentum of the binary
system $({\bm q},{\bm p})$. For a non spinning BBH system, it has been
shown that its dynamics, up to the 3PN level, can be described by the
following EOB Hamiltonian (in polar coordinates, within the plane of
the motion): 
\begin{equation}
\label{eq:Heob}
H_{\rm EOB}(r,p_{r_*},p_\varphi) \equiv M c^2\sqrt{1+2\nu (\hat{H}_{\rm eff}-1)} \  ,
\end{equation}
where
\begin{equation}
\label{eq:Heff}
\hat{H}_{\rm eff} \equiv \sqrt{p_{r_*}^2 + A(r) \left( 1 +
  \frac{p_{\varphi}^2}{r^2} 
+ z_3 \, \frac{p_{r_*}^4}{r^2} \right)} \, .
\end{equation}
Here $M\equiv M_A + M_B$ is the total mass, $\nu \equiv M_A \, M_B /
(M_A + M_B)^2$ is the symmetric mass ratio, and $z_3 \equiv 2\nu
(4-3\nu)$. In addition, we are using rescaled dimensionless
(effective) variables, namely $r \equiv r_{AB}c^2 / GM$ and
$p_{\varphi} \equiv P_{\varphi}c / (GM_A M_B)$, and $p_{r_*}$ is
canonically conjugated to a ``tortoise'' modification of
$r$~\cite{Damour:2009ic}.

A remarkable feature of the EOB formalism is that the complicated,
original 3PN Hamiltonian (which contains many corrections to the basic
Newtonian Hamiltonian $\frac{1}{2} \, {\bm p}^2 - 1/r$) can be
replaced by the simple structure (\ref{eq:Heob})-(\ref{eq:Heff}),
whose two crucial ingredients are: (i) a ``double square-root''
structure $H_{\rm EOB} \sim \sqrt{1+\sqrt{{\bm p}^2 + \cdots}}$ and
(ii) the ``condensation'' of most of the nonlinear relativistic
gravitational interactions in one function of the (EOB) radial
variable: the basic ``radial potential'' $A(r)$. The
structure of the function $A(r)$ is rather simple at 3PN,
being given by
\begin{equation}
\label{eq:3.3}
A^{\rm 3PN} (r) = 1-2u+2 \, \nu \, u^3 + a_4 \, \nu \, u^4 \, ,
\end{equation}
where $a_4 = 94/3 - (41/32)\pi^2$, and $u \equiv 1/r =
GM/(c^2r_{AB})$\label{u=1/r}. It was recently found that an excellent description
of the dynamics of BBH systems is obtained~\cite{Damour:2009kr} by:
\textit{(i)} augmenting the presently computed terms in the PN
expansion (\ref{eq:3.3}) by additional 4PN and 5PN terms;
\textit{(ii)} Pad\'e-resumming the corresponding 5PN ``Taylor''
expansion of the $A$ function. In other words, the BBH (or ``point
mass'') dynamics is well described by a function of the form
\begin{equation}
\label{eq:3.4}
A^0(r) = P^1_5\left[1-2u+2\nu u^3 + a_4 \nu u^4 + a_5\nu u^5 + 
a_6\nu u^6\right]  ,
\end{equation}
where $P^n_m$ denotes an $(n,m)$ Pad\'e approximant. It was found in
Ref.~\cite{Damour:2009kr} that a good agreement between EOB and
numerical-relativity BBH waveforms is obtained in an extended
``banana-like'' region in the $(a_5,a_6)$ plane approximately spanning
the interval between the points $(a_5,a_6)=(0,-20)$ and
$(a_5,a_6)=(-36,+520)$. In this work we will select the values
$a_5=-6.37$, $a_6=+50$, which lie within this region (we have checked
that the use of other values within the ``good BBH fit'' region would
have no measurable influence on our discussion below).

The proposal of Ref.~\cite{Damour:2009wj} for including dynamical
tidal effects in the conservative part of the dynamics consists in
simply using Eqs.~\eqref{eq:Heob}-\eqref{eq:Heff} with the following
tidally-augmented radial potential 
\begin{equation}
\label{eq:A}
A(u) = A^0(u) + A^{\rm tidal} (u)  .
\end{equation}
Here $A^0(u)$ is the point-mass potential defined in
Eq.~\eqref{eq:3.4}, while $A^{\rm tidal}(u)$ is a supplementary
``tidal contribution'' of the form
\begin{align}
\label{eq:T9}
A^{\rm tidal}=\sum_{\ell\geq 2} - \kappa_\ell^T u^{2\ell+2}\hat{A}^{\rm tidal}_\ell(u)\,,
\end{align}
where the terms $\kappa^T_\ell u^{2\ell +2}$ represent the
leading-order (LO) tidal interaction, \ie the Newtonian order tidal interaction. The dynamical
EOB tidal coefficients  $\kappa^T_\ell$ are  functions of the
two masses $M_A$ and $M_B$, of the two compactnesses ${\cal
  C}_{A,B}=GM_{A,B}/R_{A,B}$, and of the two (relativistic) Love
numbers $k_\ell^{A,B}$ of the two objects~\cite{Damour:2009,
  Binnington:2009bb, Hinderer09}:
\begin{eqnarray}
\label{eq:def_kT}
\kappa_{\ell}^{ T} &=& 2 \dfrac{M_B \, M_A^{2\l}}{(M_A + M_B)^{2\l + 1}}
\dfrac{k_\l^A}{{\cal C}_A^{2\ell + 1}} +  
\left\{ ~_A~\leftrightarrow~_B\right\} \nonumber \\
&=&\dfrac{1}{2^{2\l-1}}\dfrac{k_\l}{{\cal C}^{2\ell + 1}}
\,,
\end{eqnarray}
where the second line refers to an equal-mass binary, as the ones
considered here. Note in Table I the rather large numerical values
for the $\ell=2$ tidal coefficients: $\kappa^T_2(\C=0.12) \simeq 496$
and $\kappa^T_2(\C=0.14) \simeq 184$. In our EOB modelling we also
use the higher multipolar tidal coefficients $\kappa^T_3$ and
$\kappa^T_4$, which are even larger than $\kappa^T_2$ (\eg
~$\kappa^T_4(\C=0.12) \simeq 20318$), although their effect is marginal in
view of the higher power of $u$ (namely $u^{2\ell+2}$) with which they enter
the $A(r)$ potential.

The additional factor $\hat{A}^{\rm tidal}_\ell(u)$ in Eq.~\eqref{eq:T9} 
represents the effect of higher-order relativistic contributions to the
dynamical tidal interactions: next-to-leading--order (NLO) contributions, 
next-to-next-to-leading--order (NNLO) contributions, etc.
Here we will consider a ``Taylor-expanded'' expression
\begin{equation}
\label{eq:linear2PN}
\hat{A}^{\rm tidal}_\ell(u)=1+\bar{\alpha}_1^{(\ell)} u + \bar{\alpha}_2^{(\ell)} u^2 \ ,
\end{equation}
where $\bar{\alpha}_n^{(\ell)}$ are functions of $M_A$, ${\cal C}_A$,
and $k_\ell^A$ for a general binary. The analytical value of the
($\ell=2$) 1PN coefficient $\bar{\alpha}_1^{(2)}$ has been reported
in~\cite{Damour:2009wj} (and recently confirmed
in~\cite{Vines:2010ca}). In the equal-mass case, it yields
$\bar{\alpha}^{(2)}_1=1.25$. By contrast, there are no analytical
calculations available for $\bar{\alpha}_1^{(\ell)}$ with $\ell>2$,
nor for the 2PN tidal coefficients $\bar{\alpha}_2^{(\ell)}$. 
Indeed, one of the main aims of the present work will be to constrain 
the value of $\bar{\alpha}_2^{(2)}$ by comparing the EOB predictions 
to numerical data.

\subsection{Effective-one-body description of the waveform and radiation reaction}
\label{sbsc:eob2}

Let us first recall that the EOB formalism defines the radiation reaction from the
angular-momentum flux computed from the waveform. Concerning the waveform,
in the case of BBH systems, the EOB formalism replaces the PN-expanded
multipolar (metric) waveform $h_\lm^{\rm PN}$ by a specifically resummed 
``factorized waveform''~\cite{Damour:2007yf,Damour:2008gu}, say
$h^0_{\lm}$ (where the superscript $0$ is added to signal the absence of tidal effects).
This tidal-free multipolar waveform $h^0_{\lm}$ includes resummed versions of very high-order 
PN effects in the phase and the modulus, in particular {\it tail effects}. Actually, in the
present work, we have used a factorized waveform which includes in the modulus (but not in the phase) 
the new (5PN accurate) 
$\nu=0$ terms recently computed in~\cite{Fujita:2010xj}\footnote{
As in Ref.~\cite{Damour:2008gu} we resum the $\ell=2, m=2$ modulus by using the
Pad\'e-resummed function $f_{22}^{\rm Pf}(x;\nu)=P^3_2[f_{22}^{\rm Taylor}(x;\nu)]$.}.
 We also included in $h^0_{\lm}$
the two next-to-quasi-circular parameters $(a_1,a_2)$ as in Ref.~\cite{Damour:2009kr}\footnote{ 
Since both {\tt M2.9C.12} and {\tt M3.2C.14} are equal-mass binaries, we fix $a_1=
  -0.0439$ and $a_2=1.3077$, according to the EOB/NR comparison (for a BBH equal-mass system) 
of Ref.~\cite{Damour:2009kr}.}.

When considering tidally interacting binary systems, one needs to augment the
BBH waveform  $h_{\lm}^0$ by tidal contributions. Similarly to the additive tidal
modification \eqref{eq:A} of the $A$ potential, we will here consider 
an {\it additive} modification of the waveform, having the structure
\begin{equation}
\label{eq:4.1}
h_{\lm} =  h_{\lm}^0  + h_{\ell m}^{\rm tidal}\,.
\end{equation}
This is slightly different from the factorized form introduced in Eq.~(71) of~\cite{Damour:2009wj}
and used in~\cite{Baiotti:2010}. The above additive form turns out to be more convenient 
for incorporating higher-order relativistic corrections to the tidal waveform. 
Using the recent computation~\cite{Vines2011} of the
1PN-accurate Blanchet-Damour mass quadrupole moment \cite{Blanchet:1989ki} 
of a tidally interacting binary system (together with the Newtonian-accurate spin 
quadrupole and mass octupole) and transforming their symmetric-trace-free 
tensorial results into our $\lm$-multipolar form, we have computed the corresponding
1PN-accurate value\footnote{We leave a detailed presentation of our results to future work. Let
us however mention that, notwithstanding some statements in footnote 4 of \cite{Vines2011}, 
the 1PN-accurate (circular) quadrupolar waveform exactly 
matches the form given in Eq. (71) of \cite{Damour:2009wj} 
(which was expressed in terms of frequency-related gauge-invariant
quantities).} of $h_{22}^{\rm tidal}$, as well as the 0PN-accurate values of
$h_{21}^{\rm tidal}$, $h_{33}^{\rm tidal}$, and $h_{31}^{\rm tidal}$. 
In addition, using the general analysis of tail effects in 
Refs.~\cite{Blanchet:1992br,Blanchet:1995fr} and the 
resummation of tails introduced in 
Refs.~\cite{DamourNagar:07b,Damour:2007yf},
we were able to further improve the accuracy of these 
waveforms by incorporating (in a resummed manner) the effect 
of tails (to all orders in $M$). From a PN point of view, this 
means, in particular, that the  tidal contribution we use to the
total metric waveform is 1.5PN accurate. 
 
In summary, the EOB tidal model that we use here is analytically
complete at the 1.5 PN level. In addition, we adopt the simplifying 
assumption that the higher--multipolar tidal--amplification factors
$\hat{A}^{\rm tidal}_\ell(u)$, for $\ell>2$, are taken to coincide with
the $\ell=2$ one. This means that the EOB model that we will use
here  contains {\it only one} (yet undetermined) higher-order 
flexibility parameter, say $\bar{\alpha}_2$, that is taken to replace the 
various  $\bar{\alpha}_2^{(\ell)}$, with $\ell=\{2,3,4,\dots\}$, 
entering Eq.~\eqref{eq:linear2PN}, i.e. $\bar{\alpha}_2^{(\ell)}\equiv \bar{\alpha}_2$
(and, similarly, $\bar{\alpha}_1^{(\ell)}=\bar{\alpha}_1^{(2)}\equiv \bar{\alpha}_1$).
Note that, although this parameter is formally of 2PN order, it is
used here as an {\it effective} parametrization of all the
higher-order effects not covered by the current analytical knowledge
(both in the conservative dynamics and in the radiation
reaction). Note also that, while in the general case such a parameter
should be allowed to depend on the mass ratio and the compactnesses,
in the equal-mass case that we consider here, it is a pure number. 
We will use below the comparison between NR simulations and EOB
predictions to constrain the value of the effective higher-order
parameter $\bar{\alpha}_2$.

\subsection{PN-expanded Taylor-T4}
\label{sbsc:T4}
Tidal effects can be accounted for also via modifications of one of
the non-resummed PN description of the dynamics of inspiralling
binaries~\cite{Flanagan08, Read:2009b,
  Hinderer09}. Reference~\cite{Hinderer09}, in particular, has
recently suggested to use as baseline a time-domain T4-type
incorporation of tidal effects. We recall that the phasing of the T4
approximant is defined by the following equations
\begin{align}
\frac{d\phi_{22}^{\rm T4}}{dt} &= 2 \, x^{3/2}, \nonumber\\
\label{eq:T4bis}
\frac{dx}{dt} &= \frac{64}{5} \, \nu \, x^5 \,\left\{ a_{3.5}^{\rm Taylor} (x)
+  a^{\rm tidal}(x)\right\}\,,
\end{align}
where $a_{3.5}^{\rm Taylor}$ is the PN expanded expression describing
point-mass contributions, given by
\begin{align}
&a_{3.5}^{\rm Taylor} (x) = 1-\left(\dfrac{743}{336}+\dfrac{11}{4}\nu\right)x +4\pi x^{3/2}\nonumber\\
&+\left(\dfrac{34103}{18144} + \dfrac{13661}{2016}\nu +
\dfrac{59}{18}\nu^2\right)x^2 -\left(\dfrac{4159}{672}+\dfrac{189}{8}\nu\right)\pi x^{5/2}\nonumber\\
&+\bigg[\dfrac{16447322263}{139708800} -\dfrac{1712}{105}\gamma -\dfrac{56198689}{217728}\nu +\dfrac{541}{896}\nu^2 \nonumber\\
&-\dfrac{5605}{2592}\nu^3 +
  \dfrac{\pi^2}{48}(256+451\nu)-\dfrac{856}{105}\ln(16x)\bigg]x^3\nonumber\\
&+\left(-\dfrac{4415}{4032} + \dfrac{358675}{6048}\nu + \dfrac{91495}{1512}\nu^2\right)\pi x^{7/2}
\end{align}
and where $a^{\rm tidal}$ is the tidal contribution. From~\cite{Vines2011} 
the latter is given at 1PN accuracy by
\begin{align}
a^{\rm tidal}(x)&=\sum_{I=A,B}a_{\rm LO}(X_I) x^5(1 + a_1(X_I) x)\ ,
\end{align}
with
\be
a_{\rm LO}(X_I) = 4 \hat{k}^I_2\dfrac{12-11X_I}{X_I}
\ee
and
\be
\label{eq:a1AB}
a_1(X)=\dfrac{4421 - 12263 X + 26502 X^2 - 18508 X^3}{336 (12-11 X)},
\ee
where we introduced the auxiliary quantity
\be
\hat{k}_2^I \equiv k_2^I \left(\dfrac{X_I}{\C_I}\right)^5\qquad I=A,B .
\ee
In the particular case of equal-mass binaries, $X_A=X_B=X=1/2$, 
$\C_A=\C_B=\C$, and the tidal contribution $a^{\rm tidal}(x)$ has the form
\begin{equation}
\label{t4:lo}
a^{\rm tidal}(x)=26 \, \k_2^T x^5 \, ( 1 + a_1^{\rm T4} x ),
\end{equation}
with $a_1^{\rm T4}=5203/4368 \approx 1.19$.

Similarly to the inclusion of yet undetermined  higher-order effects in the
tidally-augmented EOB formalism via the effective parameter $\bar{\alpha}_2$,
we will consider below an {\it effective} modification of the 1PN result (\ref{t4:lo})
of the form
\be
\label{eq:T4PN_factor}
{a}^{\rm tidal}(x)=26 \, \k_2^T x^5 \,( 1 + a_1^{\rm T4} x + a_2^{\rm T4}x^2),
\ee
with an effective higher-order parameter\footnote{We found that the 1.5PN fractional contribution
$a_{3/2}^{\rm T4} x^{3/2}$ to $a^{\rm tidal}(x)$, predicted by our 1.5PN-accurate EOB waveform,
has (like the 1PN contribution) only a small effect on the phasing compared to the
large amplification that we will need to agree with NR data. This is why we only consider here,
for simplicity and for easier comparison with the 2PN EOB parameter $\bar{\alpha}_2$, the formally
2PN parameter $a_2^{\rm T4}$.} $a_2^{\rm T4}$, which we will constrain by comparing
NR data to the T4-predicted phasing.

Let us mention that, in the case of inspiralling BBH systems, several 
studies~\cite{Damour:2007yf,Boyle:2008ge,Damour:2008te} have shown 
that the nonresummed Taylor-T4 description of the GW phasing was 
significantly less accurate than the EOB description, especially 
for mass ratios different from one. Ref.~\cite{Damour:2009wj}
has also shown that, in the presence of tidal effects, it was predicting 
GW phases that differed by more than a radian with respect to the 
tidal-completed EOB model. Below, we will investigate how the T4 phasing 
based on Eq.~\eqref{eq:T4bis} differs from the EOB one, both in the absence
(Eq.~\eqref{t4:lo}) and in the presence (Eq.~\eqref{eq:T4PN_factor}) of
the higher-order parameter $a_2^{\rm T4}$.

\section{Characterizing the phasing: the $Q_{\omega}(\omega)$ function}
\label{sec:Qomega}

In order to measure the influence of tidal effects it is useful to
consider the ``phase acceleration''\footnote{In the text of this Section
$t$ and $\omega$ denote the dimensionless quantities $\hat{t}\equiv t/M$ and $\hat{\omega}\equiv M\omega$.} 
$\dot\omega \equiv d \, \omega / dt \equiv d^2 \phi / dt^2$ as a function of $\omega$, say 
$\dot{\omega}=\alpha(\omega)$ (here $\omega\equiv \omega_{22}$ 
can be either the curvature or the metric 
instantaneous GW frequency). 
Indeed, as emphasized in~\cite{Damour:2007yf}, the function $\alpha(\omega)$
is independent of the two ``shift ambiguities'' that affect the 
GW phase $\phi (t)$, namely the shifts in time and phase. 
The $\alpha(\omega)$ diagnostics (especially in its Newton-reduced
form $a_\omega = \alpha(\omega)/(c_\nu \omega^{11/3})$, with
$c_{\nu} = \frac{12}{5} 2^{1/3}\nu$, is a useful intrinsic measure
of the quality of the waveform and it has been used extensively
in recent analyses of BBHs~\cite{DamourNagar:07b, Damour:2008te,Damour:2010, Bernuzzi:2010ty}.

Here we will use another dimensionless measure of the
phase acceleration: the function $Q_{\omega}(\omega)$. It is defined as the
derivative of the (time-domain) phase with respect to the logarithm of the (time-domain) frequency:
\begin{equation}
\label{eq:5.15}
Q_{\omega} (\omega) = \frac{d\phi}{d \, \ln \, \omega} =
\frac{\omega \, d \phi / dt}{d\omega / dt} =
\frac{\omega^2}{\dot\omega} = \frac{\omega^2}{\alpha(\omega)} \, .
\end{equation}
Note that, as a consequence of this definition, the 
(time-domain) GW phase $\phi_{(\omega_1,\omega_2)}$ 
accumulated between frequencies
$(\omega_1,\omega_2)$ is given by the following integral: \be
\label{eq:phi_w1_w2}
\phi_{(\omega_1,\omega_2)}=\int_{\omega_1}^{\omega_2}Q_\omega d\ln\omega\,.
\end{equation}

\begin{figure}[t]
\begin{center}
\includegraphics[width=0.45\textwidth]{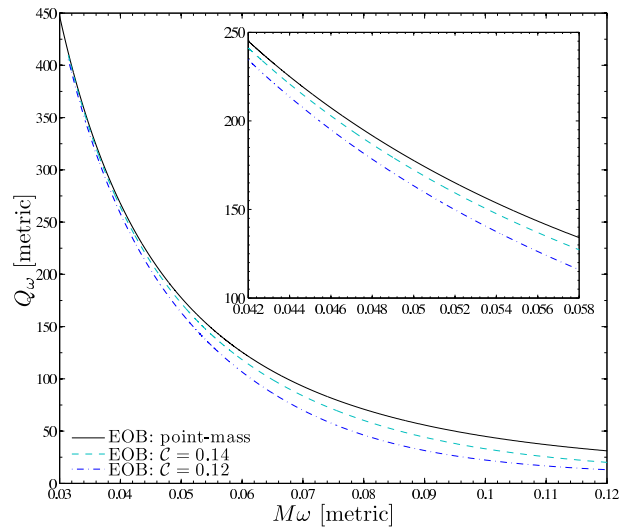}
\caption{\label{fig:explaining_Q_omega}Exploring the properties of
  $Q_\omega$ curves computed within the EOB model for three binary
  systems. Tidal interactions are approximated at LO. The inset shows a magnification, in order to
  highlight the differences among the curves.}
  \end{center}
\end{figure}
\begin{figure*}[t]
\begin{center}
\includegraphics[width=0.45\textwidth]{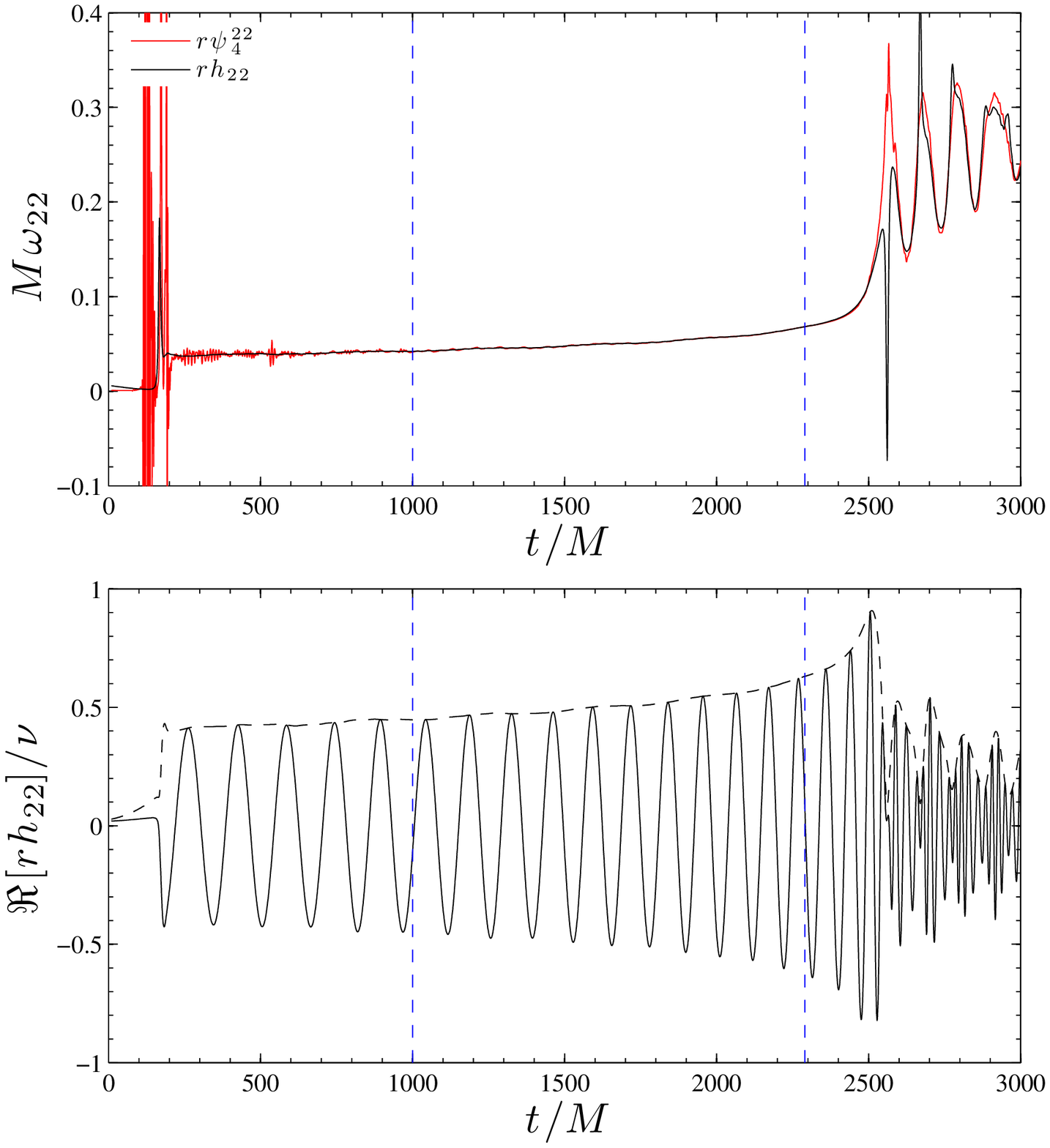}
\includegraphics[width=0.45\textwidth]{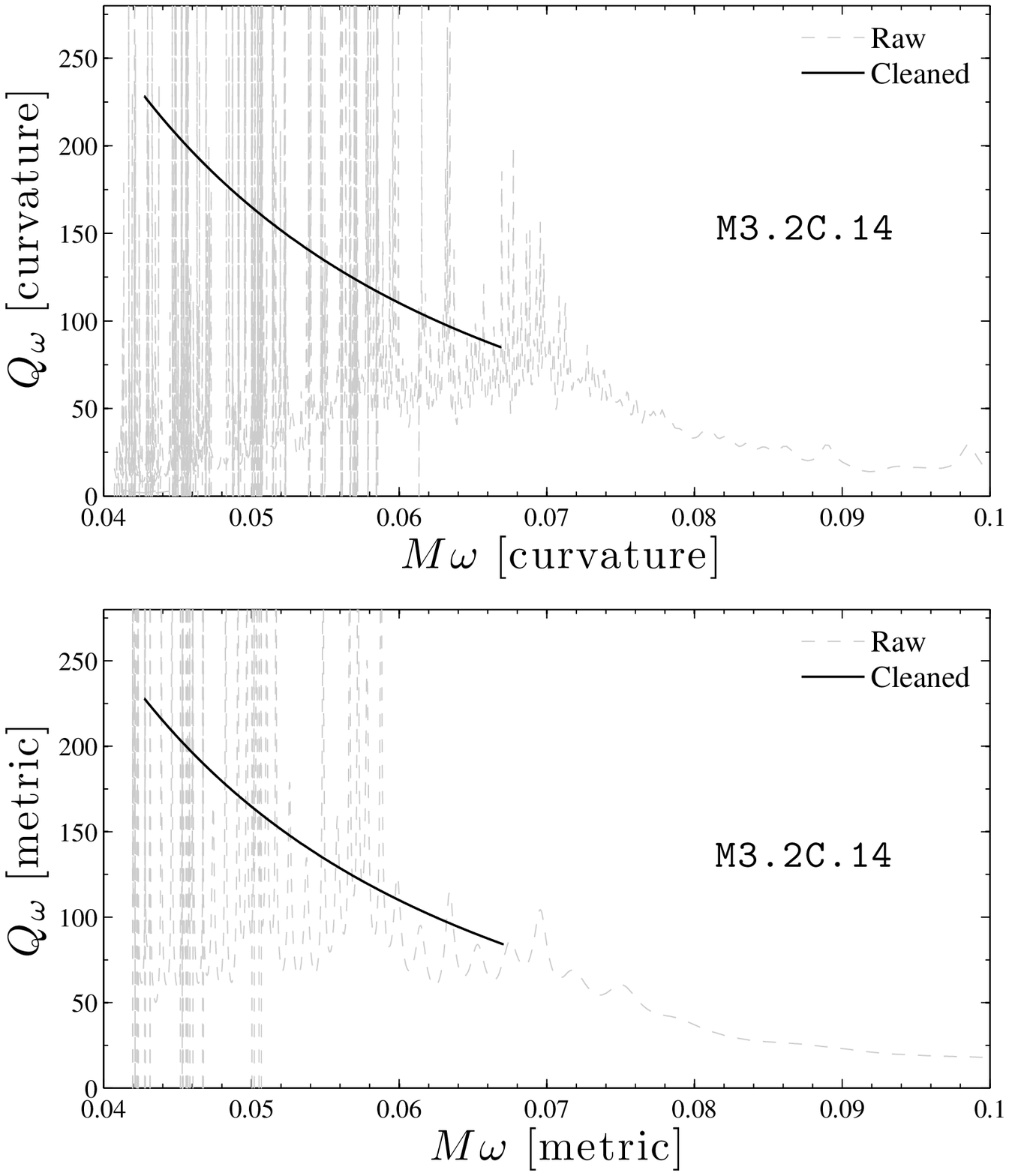}
\caption{\label{fig:cleaning} Obtaining the $Q_\omega$ diagnostic 
from a suitable fitting procedure of the GW phase (for both
curvature and metric waveforms). The two vertical lines on the 
left panels indicate the time interval $\Delta t/M=[1000,2290]$ 
where we fit the NR phase with Eq.~\eqref{eq:phi}. For completeness 
we also display the real part of the metric waveform. 
On the right panels, the (light) dashed lines refer to the $Q_\omega$ 
obtained by direct numerical differentiation of the raw data;
the solid lines are instead obtained from the fitted phase.
Although the curves displayed here refer to model~\texttt{M3.2C.14},
similar results are  obtained also for the binary~\texttt{M2.9C.12}.}
  \end{center}
\end{figure*}

Stated differently, the function $Q_{\omega} (\omega)$ measures the
number of GW cycles spent by the binary system within an octave of the
GW frequency $\omega$ (it is therefore analogous to the ``quality
factor'' $Q$ of a damped oscillator). 
Let us also note that, in the
stationary phase approximation, $Q_{\omega}$ enters as an
amplification factor of the signal, so that the squared
signal-to-noise ratio is equal to~\cite{Damour:2000gg}
\begin{equation}
\label{eq:5.16}
\rho^2 = 4 \int d \ln \, \omega \ \frac{Q_{\omega} (\omega) \, 
A^2 (\omega)}{\omega \, S_n (f)}\,,
\end{equation}
where $A$ denotes the amplitude of the time-domain metric waveform
and where $S_n (f)$ denotes the one-sided noise power spectral density and
$f \equiv \omega /(2 \pi)$.

In view of its definition, $Q_{\omega}$ is a useful {\it quantitative
  indicator} of the physics driving the variation of $\omega$.
Indeed, a change of $Q_{\omega}(\omega)$ of the order $\pm 1$ during a
frequency ``octave'' $\ln(\omega_2/\omega_1)=1$ corresponds to a local
dephasing (around $\omega$) of $\Delta \phi \simeq \pm 1$~rad. Because
such a dephasing (if it occurs within the sensitivity band of the
detector) can be expected to significantly affect the measurability of
the signal, it is probably necessary to model $Q_\omega$ with an
absolute accuracy of about $\pm 1$ (see Ref.~\cite{Damour:2010} for a
quantitative discussion of the admissible error level on $Q_\omega$ in
the BBH context).

We start our analysis by comparing the $Q_\omega$ functions (as
predicted by the EOB formalism) for the (metric) gravitational
waveforms $h_{22}$ generated by three (equal-mass) binary models,
namely a BBH and the two BNS systems discussed in
Sec.~\ref{numsetup}. To simplify the discussion, these functions are
computed with the LO tidal interaction $\hat{A}_\ell(u)=1$. [We will
  separately study below the effect of changing $\hat{A}_\ell(u)$.]

Figure~\ref{fig:explaining_Q_omega} compares the properties of the
$Q_\omega$ functions by showing together the curves for the three
binaries versus their corresponding GW frequency. A number of remarks
are worth making. First, $Q_\omega$ is a large number that diverges in
the small-frequency limit. This follows from the fact that in the
limit $\omega\to 0$ one has $\alpha(\omega) \sim c_\nu \omega^{11/3}$, and
then, via Eq.~\eqref{eq:5.15}, $Q_\omega = 1/(c_\nu
\omega^{5/3})\sim(c/v)^5$. Second, the presence of tidal interactions
{\it decreases} the ``point-mass'' value of $Q_\omega$ by an amount that is
(essentially) proportional to $\kappa_2^T$. In other words, tidal
effects ``accelerate'' the inspiral by reducing the number of cycles
spent around a given frequency. In particular, BBHs (which have vanishing
tidal constants \cite{Damour:2009, Binnington:2009bb}) are effectively the binaries that spend
the largest time at any given frequency. Finally, note that since $Q_\omega$ is
a large number, the fact that the curves look relatively close on the
large-scale plot can be misleading, since the corresponding accumulated
relative phase difference can actually be large (see inset, which shows that the 
absolute differences between the various $Q_\omega(\omega)$ is of order $10$,
corresponding to integrated dephasings of order $10$~radians). 

Although the calculation of the phase ``quality-factor'' $Q_\omega$ is
straightforward within the EOB framework, this is not the case when
$Q_\omega$ is to be calculated from the NR (either curvature or
metric) waveforms. Indeed, the direct computation of the $Q_\omega$
functions from raw data is in general made difficult by the presence
of both high-frequency noise in $\omega(t)$ and of low-frequency
oscillations probably due to a residual eccentricity. This is
illustrated in the right panels of Fig.~\ref{fig:cleaning}, where we
show with (light) dashed lines the raw NR $Q_\omega$ functions
obtained by direct time-differentiation of the NR curvature (top
panel) or metric (bottom panel) phase for the
binary~\texttt{M3.2C.14}. A fourth order accurate finite differencing
algorithm has been used to compute the derivatives. Similar results
have been obtained also for the binary~\texttt{M2.9C.12}.

The right panel of Fig.~\ref{fig:cleaning} shows that the two
time-derivatives involved in the definition of $Q_\omega(\omega)$
amplify considerably the high-frequency noise contained in the NR
phase evolution, and make it impossible to extract a reliable value of
$Q_\omega(\omega)$ from such a {\it direct} numerical attack. To
tackle this problem, one needs to filter out the high-frequency
numerical errors in the time-domain phase before effecting any
time-differentiation. To do this, we found useful to ``clean'' the
phase $\phi(t)$ by fitting the NR phase to an analytic expression that
is inspired by the PN expansion. More precisely, after introducing a
formal ``coalescence'' time $t_c$ and defining the quantity
\begin{equation}
\label{eq:def_x}
x\equiv\left[\dfrac{\nu}{5}(t_c -t)\right]^{-1/8}\,,
\end{equation}
we  fitted the time-domain NR phase $\phi^{\rm NR}(t)$  to an expression of the form
\begin{align}
\phi(t; t_c, p_2, p_3, p_4, \phi_0)  &= \phi_0 + -\dfrac{2}{\nu}x^{-5}\nonumber\\
\label{eq:phi}
                                    &\times \left(1 + p_2 x^2 + p_3 x^{3} + p_4 x^4 \right)\,.
\end{align}
In this expression, we have set the lower coefficients $p_0$ and $p_1$ to
$p_0=1$ and $p_1=0$, as suggested by the corresponding lowest-order PN
expression (see, \eg ~Eq.~(234) of~\cite{Blanchet06}), but we left
$t_c$, $\phi_0$, and the higher-PN $p_i$'s as free coefficients to be
determined from the NR data. The basic idea is that of using a simple
analytical form that incorporates the leading trend of $Q_\omega$ to
remove the influence of the numerical errors while leaving some
flexibility in the subleading part of the phase evolution that is
influenced by tidal effects. We view the fitting parameters
$p_2,p_3,p_4$ as effective parameters for describing tidal-phasing
effects.

Such a fit of the phase evolution can be reliably done only in a
limited time interval. Indeed, one has to cut off both the early
phase of the inspiral (where the numerical data is too noisy), and the
last few cycles before the merger (where the PN-based fit is no longer
a good approximation). We present in Appendix~\ref{sec:cleaning} a
detailed discussion of the optimal choice of the time interval where
to make the fit, as well as a series of consistency checks. See also
the discussion at the end of Sec.~\ref{sec:errors_freq}.

Let us start by discussing the application of this procedure to the GW
phase (both curvature and metric) of the binary model
$\texttt{M3.2C.14}$. The result of this fitting is shown by the solid
lines in the right-panels of Fig.~\ref{fig:cleaning} (top, curvature
phase; bottom, metric phase). The time interval on which we could
reliably apply the fitting procedure is $ I_t/M=[1000,2290]$. This
time window is indicated by the dashed vertical lines in the top-left
panel of Fig.~\ref{fig:cleaning}, were we show together the time
evolution of both the curvature (dashed, red online) and metric
(solid) GW frequencies. For completeness, the lower-left panel of the
same figure translates this information in terms of GW cycles of the
metric waveform. Note that this time interval excludes the first $4$
GW cycles (whose NR frequency is indeed seen to be quite noisy), but
covers about $10$ GW cycles, and ends around $2$ GW cycles before
the merger (defined as the maximum of the modulus of the metric waveform;
the modulus of the metric waveform is indicated by a dashed line on
the left-bottom panel of the figure). The corresponding frequency
interval can be visualized on the right panels, and is listed in the
fifth column of Table~\ref{tab:error_freq}. Similar results are obtained
also for the {\tt M2.9C.12} data (see Fig.~\ref{fig:Qomega0_EOB}
below). In this case, the time interval we use is $I_t/M=[1300,3366]$,
with the corresponding frequencies listed in the seventh column of
Table~\ref{tab:error_freq}. Finally, notice that for this model the
inspiral is longer than in the previous case and so this interval
actually corresponds to $14$ GWs cycles. In addition, similarly to
the other case, our choice of fitting interval excludes the first
$5.5$ GW cycles, and ends about $2$ GW cycles before merger.

As we will discuss below, although the frequency windows where our
cleaning procedure allowed us to compute an estimate of the NR
$Q_\omega(\omega)$ functions do not cover the full inspiral, these
estimates will give us access to important information for performing
quantitative comparisons with the predictions of the EOB (and Taylor
T4) analytical models.

\section{Numerical Error-Budget}
\label{sec:errors}

The aim of this section is to discuss the various errors affecting
the numerical waveforms extracted (for both models) at  $500\Msun$
and computed with the highest resolution. 
Such a discussion  will in turn allow us to estimate an uncertainty range
on  the analytical parameter $\bar{\alpha}_2$ representing the 
not-yet-calculated, high-PN-order tidal effects entering the 
EOB description of the phasing. 

We will discuss in turn the numerical errors entailed by three
different effects: (i) the choice of EOS (isentropic versus
non-isentropic evolution); (ii) the finite extraction radius; (iii)
the finite resolution. We will perform this analysis both by
comparing waveforms in the time domain and by means of the $Q_\omega$
diagnostic.

\subsection{Time-domain analysis}
\subsubsection{Non-isentropic evolutions}
\begin{figure}[t]
\begin{center}
\includegraphics[width=0.45\textwidth]{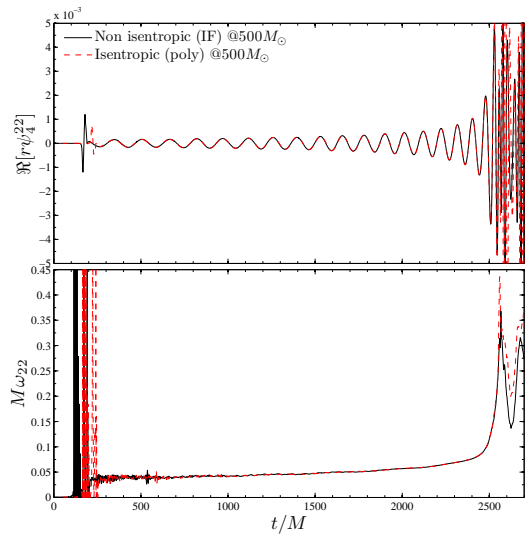}
\caption{\label{fig:IF_poly}Comparing waveforms from
  isentropic (dashed) and non-isentropic (solid) evolution for BNS
  model \texttt{M3.2C.14}. Waveforms are computed with the highest
  resolution and extracted at $r_{\rm obs}=500\Msun$. The
  corresponding phase difference $\phi^{\rm poly_{HR}500}-\phi^{\rm
    IF_{HR}500}$ is displayed in Fig.~\ref{fig:error_time}.}
  \end{center}
\end{figure}
\begin{figure}[t]
\begin{center}
 \includegraphics[width=0.45\textwidth]{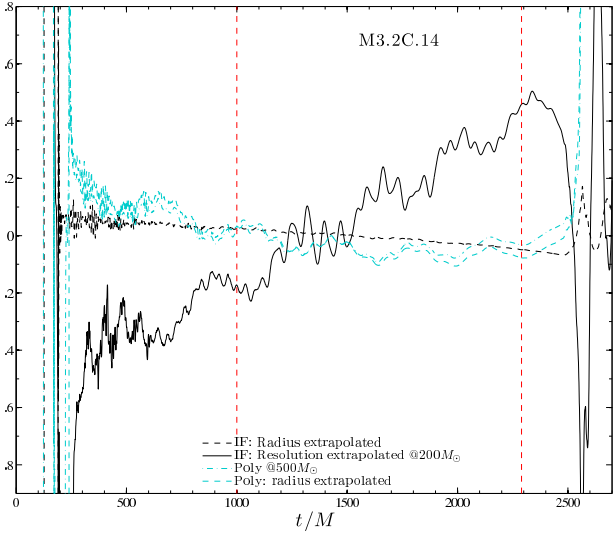}\\
\vspace{10 mm}
\includegraphics[width=0.45\textwidth]{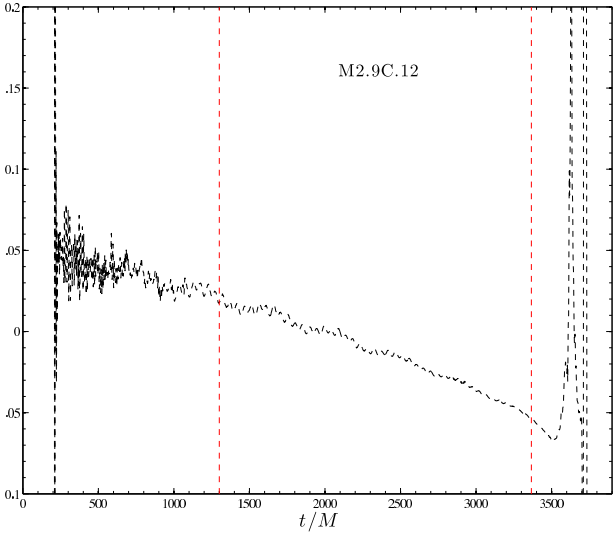}
\caption{\label{fig:error_time}Estimate of the phase uncertainty in
  the time domain for model~\texttt{M3.2C.14} (top)
  and~\texttt{M2.9C.12} (bottom). The figure shows the phase
  difference between different ``post-processed'' numerical curvature
  waveforms $r\psi_4^{22}$ (in particular, extrapolated in resolution
  and/or extraction radius) and the one obtained with the ideal-fluid EOS and
  extracted at $r_{\rm obs}=500\Msun$.}
\end{center}
\end{figure}

As discussed in Sec.~\ref{numsetup}, we have evolved the binaries
using either a (isentropic) polytropic EOS or a (non-isentropic)
ideal-fluid EOS. We recall that, in the absence of large-scale shocks
(like those taking place at the merger), the two EOSs are equivalent
and should therefore yield evolutions that differ only at machine
precision. In practice, however, when using the ideal-fluid EOS small
shocks are produced in the very low-density layers of the stars even when
these orbit~\cite{Baiotti08}. These small shocks channel some of the
orbital kinetic energy into internal energy, leading to small
ejections of matter (\ie amounting to a total of $\sim 10^{-6}
M_{\odot}$), and are thus responsible for slight differences even
during the inspiral. Since we are here presenting the results of
simulations that are considerably longer than any presented so far and
in particular of those in Refs.~\cite{Baiotti08,Baiotti:2009gk}, it is
important to quantify the influence of these non-isentropic
effects. Concentrating on model \texttt{M3.2C.14}, we show in the
top-panel of Fig.~\ref{fig:IF_poly} the real parts of the
$r\psi_4^{22}$ waveforms computed with the two EOSs as extracted at
$r_{\rm obs}=500M_\odot=165.1 M$. The bottom panel displays the
corresponding instantaneous frequencies for completeness. As
customary in comparing waveforms in the time domain, one allows for
arbitrary relative time and phase shifts $(\tau,\alpha)$. These
quantities can be determined in various ways, for example by means of
the two-frequency pinching technique of Ref.~\cite{Damour:2007vq}. In
this paper we find it useful to use the method used in
Ref.~\cite{Boyle:2008ge} to compute $(\tau,\alpha)$. More precisely,
given two timeseries of the phase $\{\phi_1(t_i),\phi_2(t_i)\}$
defined on a given time interval $[t_L,t_R]$ that is covered by $N$
numerical points $t_i$, with $i=1,2,\dots,N$, we define the quantity
\be \Delta\phi(t_i,\tau, \alpha) = \phi_2(t_i+\tau) - \phi_1(t_i)
-\alpha \ee and determine $\tau$ and $\alpha$ such that they minimize
the ``reduced'' $\chi^2$ quantity
\begin{align}
\label{eq:chi2}
\widehat{\chi}^2 = \dfrac{1}{N}\sum_{i=1}^N (\Delta\phi(t_i,\tau,\alpha) )^2.
\end{align}
The minimization on $\alpha$ is done analytically, while that on $\tau$
is done numerically. Note in addition that the square root of the minimum
value of Eq.~\eqref{eq:chi2}, say
\begin{equation}
\sigma_{\Delta\phi} = \sqrt{\dfrac{1}{N}\sum_{i=1}^N (\Delta\phi(t_i,\tau,\alpha))^2_{\rm min}}
\end{equation}
has the meaning of a root-mean-square deviation of the phase
difference $\Delta\phi$ over the interval $[t_L,t_R]$; as such, it can
also be employed to give a quantitative measure of a phase difference
(and thereby of some phase errors).\footnote{We note in passing that
  the alignment procedure also highlights the weak dependence on the
  EOS of the late part of the waveform: although the inspiral of the
  non-isentropic waveform is about $150M$ longer than the
  corresponding isentropic one, the growth of $M\omega_{22}$ (and the
  corresponding phasing) is qualitatively and quantitatively very
  close for both models until $M\omega_{22}$ peaks for the first
  time.}  The phase difference $\Delta\phi(t)\equiv
\phi_2(t)-\phi_1(t)=\phi^{\rm poly_{HR}500}-\phi^{\rm IF_{HR}500}$
(least-square minimized on the time interval $[t_L,t_R]/M=
[300,2540]$) is represented as a dash-dotted line (solid light
blue) in Fig.~\ref{fig:error_time}. One sees that the
instantaneous phase difference varies roughly between $+0.2$~rad
and $-0.1$~rad on this time interval, which corresponds to a
``two-sided''~\cite{Damour:2007vq} phase uncertainty of the order 
$\Delta\phi=\pm\dfrac{1}{2}(0.2-(-0.1))=\pm 0.15$
rad. The information of Fig.~\ref{fig:error_time} is completed by
Table~\ref{tab:error_time}, where we list the $\ell^{\infty}$
norm of the phase difference [\ie the maximum absolute value of
$\Delta\phi(t)$], labelled $||\Delta\phi||^{\infty}$, the
root-mean-square $\sigma_{\Delta\phi}$ as computed above, and
the corresponding time interval $[t_L,t_R]$ that is used to
compute $(\alpha,\tau)$. Note that $\sigma_{\Delta\phi}$ gives
a measure of the phase difference which is always significantly
smaller than the $\ell^{\infty}$ norm. Indeed, these two
quantities measure different aspects of a phase difference, and,
when the time variation of $\Delta\phi(t)$ is dominated by
low-frequency effects (which can be roughly modelled as power
laws), the averaging involved in the definition of
$\sigma_{\Delta\phi}$ will lead to a smallish ratio
$\sigma_{\Delta\phi}/||\Delta\phi||^{\infty} < 1$ linked to
integrals of the type $\int_0^1 dt \, t^ {2 n}= 1/(2 n+1)$.

\subsubsection{Finite-radius extraction}

We next discuss the phasing error introduced by the fact that our
high-resolution target waveforms, for both models, are extracted at
the finite coordinate radius $r_{\rm obs}=500\Msun$. Note that, when
expressed in units of the gravitational mass $M$ of the binary at
infinite separation, this value corresponds to $r_{\rm
  obs}=134.9M$ for {\tt M2.9C.12} and $r_{\rm obs} = 165.1M$ for {\tt
  M3.2C.14}, \ie, for one model waves are actually extracted slightly
farther than for the other. For both models we have at our disposal
several extraction radii, so that we can estimate the phasing error
linked to the finite extraction radius as follows: (i) We used the raw
$r\psi_4^{22}$ data extracted at radii $r=\{400, \, 450,\,
500\}\Msun$; (ii) We time-shifted them so that this triplet of
timeseries is expressed as a function of the (coordinate) retarded
time $t_*=t-r-2M_{\rm ADM}\ln\left[r/(2M_{\rm ADM})-1\right]$; (iii)
We separated each curvature waveform in phase and amplitude as
functions of $u \equiv 1/r$ (\cf page~\pageref{u=1/r}); (iv) We fitted each resulting triplet of timeseries
to a linear polynomial in the triplet of inverse extraction radii:
$c^{\infty}(t_*)+c_1(t_*)/r$. The quantities $c^{\infty}(t_*)$ [\ie
$A^{\infty}(t_*)$ and $\phi^{\infty}(t_*)$] yield estimates of the
amplitude and phase of the infinite-radius extrapolation of
$r\psi_4^{22}$. We then compare the radius-extrapolated phase
$\phi^{\infty}(t_*)$ to the phase extracted at the outermost radius,
allowing for additional time and phase shifts (which are determined by
the least-square minimization discussed above).
 
The time evolution of the phase differences computed in this way are
shown in Fig.~\ref{fig:error_time} for model~\texttt{M3.2C.14} (top
panel, dash-line) and for~\texttt{M2.9C.12} (bottom panel). This
local information is completed by the ``global'' quantitative
information ($||\Delta\phi||^{\infty}$, $\sigma_{\Delta\phi}$) listed
in the last two columns of Table~\ref{tab:error_time}. On the
basis of this analysis, we estimate that, for both models, the phase
uncertainty due to finite extraction is of order $\Delta\phi\approx
\pm 0.05$~rad almost up to the merger, \ie roughly $100\,M$ before the
peak of the GW frequency.

\begin{table}[t]
\caption{\label{tab:error_time} Uncertainty estimates on the phase (in
  radians) of $r\psi_4^{22}$, computed in the time domain, for both
  models. From left to right, the columns report: the model name, the EOS, the
  coordinate extraction radius, the type of extrapolation that is
  performed on the waveform (either in extraction radius or
  resolution), the time interval on which the $\chi^2$ of the phase
  difference is minimized, the $\ell^{\infty}$ norm of the phase
  difference over this interval, and the root-mean-square of the phase
  difference.}
 \begin{center}
  \begin{ruledtabular}
  \begin{tabular}{lclcccc}
Model&    EOS & $r_{\rm obs}$ & Extrap. & $[t_L,t_R]$ & $||\Delta\phi||^{\infty}$ & $\sigma_{\Delta\phi}$  \\
&     & $[\Msun]$ &  & $[M]$ & [rad] & [rad]  \\
   \hline \hline
{\tt M3.2C.14}&  IF   & $500$  & radius       & $[400,2650]$ &  $0.17$  &  $0.035$  \\
{\tt M3.2C.14}&  IF   & $200$  & resolution   & $[400,2650]$ &  $1.29$  &  $0.300$  \\
{\tt M3.2C.14}&  poly & $500$  & $-$         & $[300,2540]$ &  $0.21$  &  $0.057$  \\
{\tt M3.2C.14}&  poly & $500$  & radius       & $[300,2550]$ &  $0.43$  &  $0.080$  \\   
   \hline
{\tt M2.9C.12}&   IF  & $500$  & radius      &$[250,3650]$   &  $0.31$ & $0.035$      \\
 \end{tabular}
\end{ruledtabular}
\end{center}
\end{table}

\subsubsection{Finite-resolution error}
\label{sec:resolution}
Finite-resolution errors have already been discussed in detail in our
previous work~\cite{Baiotti:2009gk}, which used the same numerical setup
(\ie the same resolution and grid structure) adopted here. Skipping
the details, we recall that it was shown there that, at the resolution
that we are using in this work, the dynamics and waveforms are in a
convergent regime, with a convergence rate $\sigma$ that is $\simeq
1.8$ during the inspiral phase and drops to $\simeq 1.2$ after the
merger, when large-scale shocks appear. As the computational cost of the calculations
presented here is already at the limit of what can be reasonably
afforded, we have decided to estimate the truncation-error of our present waveform by assuming
that the inspiral convergence rate $\sigma \simeq 1.8$ found in our
previous work~\cite{Baiotti:2009gk} approximately holds in the present 
(numerically similar) case. We have then selected the more compact
binary {\tt M3.2C.14} and used only two simulations with different resolutions. 
More specifically, we have considered a ``high-resolution''  simulation, where the
finest refinement level has a resolution $\h_{\rm H}=0.12\Msun$, and a
``low-resolution''  simulation, with $\h_{\rm L}=0.15\Msun$.
For this particular comparison the waveforms are extracted at $r_{\rm obs}=200\Msun$.
When comparing the low- and high-resolution curvature waveforms,
after suitable $(\tau,\alpha)$ alignment, one discovers that 
the phase difference accumulated between the two resolutions over a
timescale of $2300M$ during the inspiral is about $0.45$~rad 
(corresponding to a relative error of $\simeq 0.36\%$).
Using the convergence rate discussed above, we can now 
Richardson-extrapolate the results obtained with the two resolutions and 
obtain an estimate of the ``infinite-resolution'' waveform. 
More precisely, we model the suitably aligned, low- and high-resolution phase 
evolutions as
\begin{align}
\phi_{\h_{\rm H}}(t) &= \phi_0(t) + k(t) (\h_{\rm H})^\sigma, \\
\phi_{\h_{\rm L}}(t) &= \phi_0(t) + k(t) (\h_{\rm L})^\sigma,
\end{align}
where $\phi_0(t)$ represents the infinite-resolution phase ($\Delta\to
0$). From the above equations, we obtain the following estimate of
the infinite-resolution extrapolation of the phase evolution
\be
\label{eq:phi_extr}
\phi_0(t) = \dfrac{(\h_{\rm L})^\sigma \phi_{\h_{\rm H}}(t) - (\h_{\rm H})^\sigma\phi_{\h_{\rm L}}(t)}
{(\h_{\rm L})^\sigma - (\h_{\rm H})^\sigma}\,.
\end{equation}

\begin{figure*}[t]
\begin{center}
\includegraphics[width=0.45\textwidth]{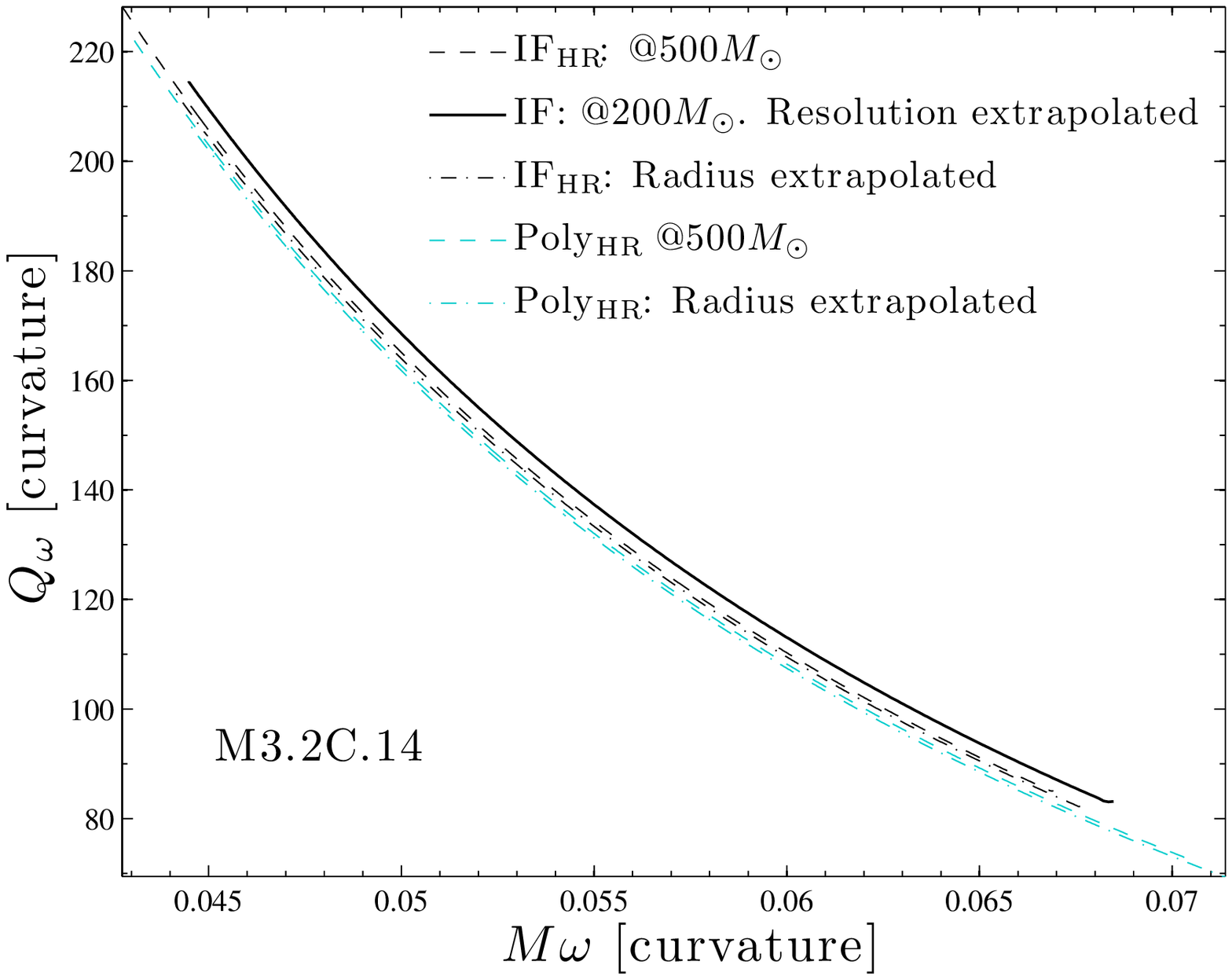}
\includegraphics[width=0.45\textwidth]{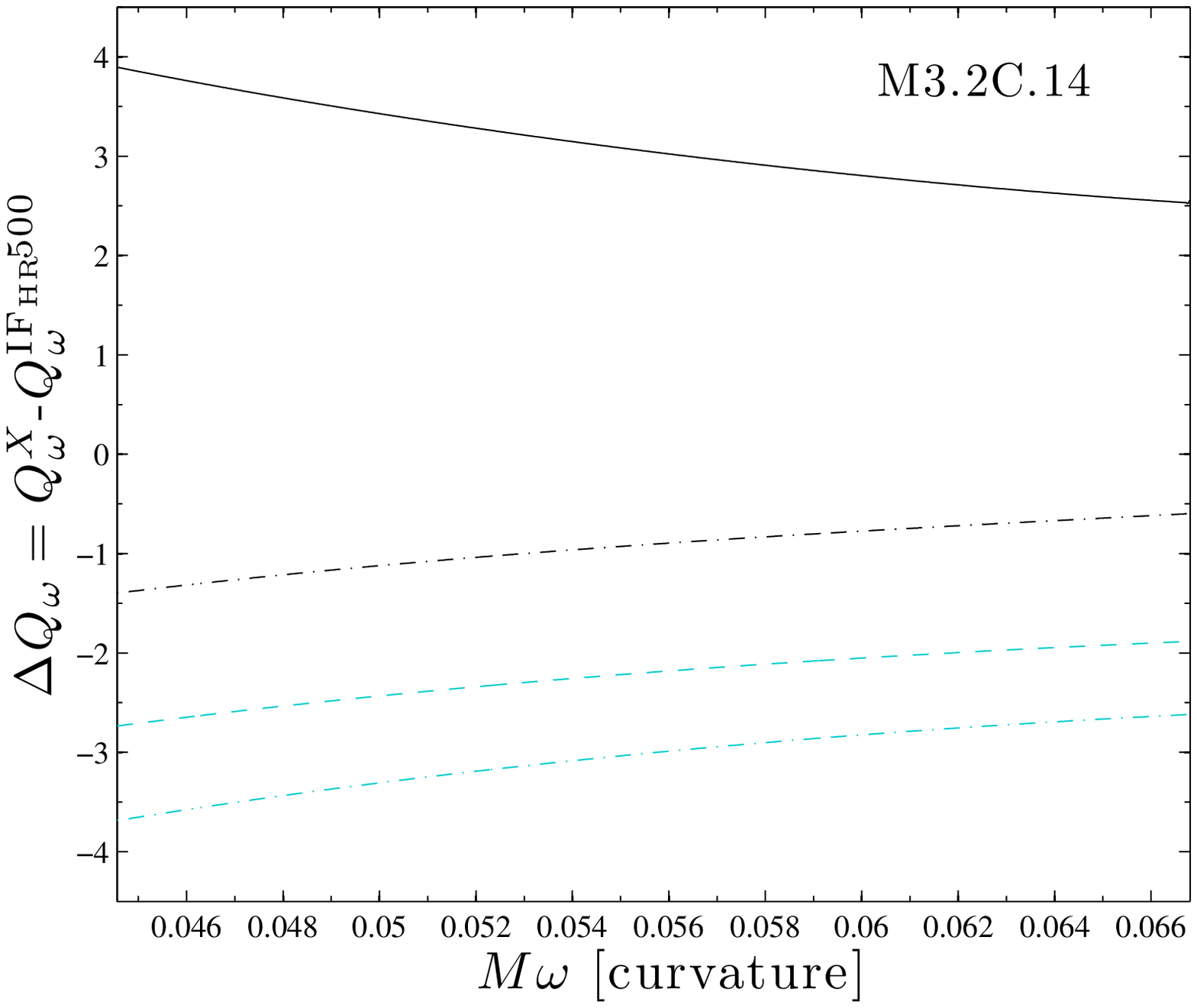}
\caption{\label{fig:error_freq}Left panel: span of $Q_\omega$'s due to the various approximations 
to the curvature waveforms from model {\tt M3.2C.14}. Right panel: the corresponding differences 
$\Delta Q_\omega= Q_\omega^X - Q_\omega^{{\rm IF_{HR}}500}$ between the  various curves and the fiducial one obtained from the 
phase computed at the highest resolution and extracted at $500\Msun$.}
\end{center}
\end{figure*}

\begin{table*}[t]
\caption{\label{tab:error_freq} Uncertainty estimates on the
  $r\psi_4^{22}$ phase of the ${\rm IF_{HR}}500$ fiducial simulations
  obtained from integration of the differences between $Q_\omega$'s.
  From left to right the columns report: the model name, the EOS, the coordinate
  extraction radius, the type of extrapolation that is performed on
  the waveform, the frequency interval $MI_\omega$ where the cleaning
  procedure is applied, the corresponding time interval $I_t$, the
  accumulated phase difference $\Delta\phi_{\psi_4}=\phi^X-\phi^{\rm
    IF_{HR}500}$ on a {\it common} frequency interval $MI_\omega^c$,
  the number of GW cycles on the same frequency interval, and the
  relative phase difference
  $\widehat{\Delta\phi_{\psi_4}}=\Delta\phi_{\psi_4}/\phi_{\psi_4}$.
  We choose the common interval of integration to be
  $MI^c_\omega=[0.045,0.067]$ for model~\texttt{M3.2C.14} and
  $MI^c_\omega=[0.037,0.054]$ for model {\tt M2.9C.12}.}
 \begin{center}
  \begin{ruledtabular}
  \begin{tabular}{lcccccccc}
Model&  EOS & $r_{\rm obs}$ & Extrap. & $MI_\omega$ & $ I_t$ &$\Delta\phi_{\psi_4}$ & $\phi_{\psi_4}$ &  $\widehat{\Delta\phi_{\psi_4}}$ \\
&  & $[\Msun]$ &  &  & [$M$] & [rad] & [$2\pi$] & [\%] \\
   \hline \hline
{\tt M3.2C.14}&  IF   & $500$  & $-$        & $[0.041,0.068]$ & $[1000,\,2290]$ & $-$      & $9.14$ & $-$    \\
{\tt M3.2C.14}&  IF   & $500$  & radius     & $[0.044,0.069]$ & $[1000,\,2130]$ &$-0.39$   & $8.99$ & $-1.61$ \\
{\tt M3.2C.14}&  IF   & $200$  & resolution & $[0.046,0.072]$ & $[1000,\,2145]$ &$~1.28$   & $9.34$ & $~2.24$ \\
{\tt M3.2C.14}&  poly & $500$  & $-$        & $[0.041,0.069]$ & $[1000,\,2290]$ &$-0.92$   & $9.07$ & $-0.69$ \\
{\tt M3.2C.14}&  poly & $500$  & radius     & $[0.044,0.072]$ & $[1000,\,2030]$ &$-1.24$   & $8.94$ & $-2.16$ \\   
   \hline
{\tt M2.9C.12}& IF & $500$  & $-$       & $[0.036,0.058]$ & $[1300,\,3366]$ &$-$     & $13.02$  & $-$   \\
{\tt M2.9C.12}& IF & $500$  & radius    & $[0.037,0.054]$ & $[1300,\,3070]$ &$-0.18$ & $13.00$  & $-0.2$ \\
 \end{tabular}
\end{ruledtabular}
\end{center}
\end{table*}

We performed the same extrapolation also on the waveform modulus, so
as to have access to the complete extrapolated curvature waveform.
The solid line in Fig.~\ref{fig:error_time} displays the phase
difference $\phi_0^{\rm IF200}-\phi^{\rm IF_{HR}500}$. This indicates
a phase uncertainty of $\Delta\phi\approx \pm 0.5$~rad on
$\phi^{\rm IF_{HR}500}$ as measured up to about $100M$ before the
maximum of $M\omega_{22}$. See Table~\ref{tab:error_time} for the
corresponding global measures of the phase uncertainty,
$||\Delta\phi||^{\infty}$ and $\sigma_{\Delta\phi}$. Note that these
uncertainty estimates are much larger than those normally computed for
binary BH simulations for the same computational costs (see, for
instance,~\cite{Pollney:2009MP-unpublished-a}). This is the natural
consequence of the smaller resolution employable here and of the
lower-order convergence that one achieves when solving the
hydrodynamics equations. Since this error is deduced only after {\it
  assuming} a certain convergence order (in addition obtained from a
slightly different numerical setup) it will be used below only to
estimate a rough uncertainty range on the value of the higher-order
EOB tidal correction parameter $\bar{\alpha}_2$. We will comment more
on this in the next Sections.

Adding in quadrature the various uncertainties computed so far to
obtain a total error bar on the phases of the ${\rm IF_ {HR}}500$ data
for the \texttt{M3.2C.14} model would give a (two-sided) time-domain
phase uncertainty $\Delta\phi \simeq \pm\sqrt{0.15^2+0.05^2}\simeq \pm
0.16$~rad, when excluding the uncertainty due to the finite
resolution, or $\Delta\phi \simeq \pm\sqrt{0.15^2+0.05^2+0.5^2} \simeq
\pm 0.52$~rad when including it. Alternatively, if we add in
quadrature the root-mean-squares of the corresponding phase errors we
find $\sigma_{\Delta\phi} \simeq \pm 0.07$~rad, when excluding the
uncertainty due to the finite resolution, and $\sigma_{\Delta\phi}
\simeq \pm 0.32$~rad when including it. Clearly the
resolution-extrapolation error is dominating the error budget. In view
of the uncertainty in estimating this source of error, we will not
directly use these time-domain phase-error levels in estimating the
uncertainties in the comparison between the EOB, T4, and NR phasings.
As we will discuss next, we prefer to express the information gathered
above on numerical errors in terms of the corresponding $Q_\omega$
curves.

\subsection{$Q_\omega$ analysis}
\label{sec:errors_freq}

\begin{figure}[t]
\begin{center}
\includegraphics[width=0.45\textwidth]{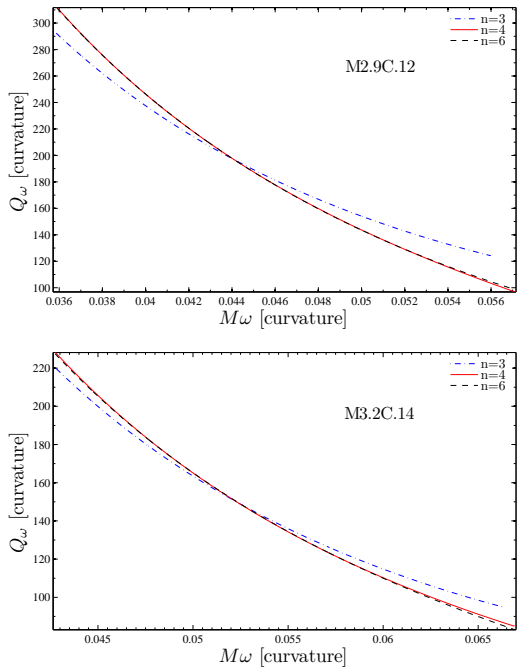}
\caption{\label{fig:Qomg_n}Sensitivity of $Q_\omega$ to the phase
  model used in the fitting procedure. Note that the $n=4$ and $n=6$
  curves are barely distinguishable on the plot. See text for
  details.}
\end{center}
\end{figure}

In Sec.~\ref{sec:Qomega} we have introduced
$Q_\omega=\omega^2/\dot{\omega}$ as a convenient, intrinsic
diagnostics to describe the phasing of the waveform. In particular, it
allows us to better visualize the influence of tidal effects on the
phasing, as well as to compute the dephasing accumulated on a given
frequency interval. It is then useful to recast the various
time-domain phase uncertainties on the high-resolution waveform
extracted at $500\Msun$ discussed above, in terms of $Q_\omega$. In
practice, we apply the cleaning procedure on each waveform of
Table~\ref{tab:error_time} so as to obtain four $Q_{\omega}$ curves.
These curves are displayed together in the left panel of
Fig.~\ref{fig:error_freq}, while the fifth column of
Table~\ref{tab:error_freq} lists the specific frequency intervals
$I_\omega$ that were selected to apply the cleaning procedure. For a
quantitative assessment of the differences between the $Q_\omega$
curves, we present in the right panel of Fig.~\ref{fig:error_freq} the
quantity $\Delta Q_\omega^X(\omega) = Q_\omega^X(\omega) -
Q_\omega^{{\rm IF_{HR}}500}(\omega)$, where the labelling $X$
indicates any curve other than our fiducial one, ${\rm IF_{HR}}500$.
Although the information conveyed by this figure is qualitatively
analogous to the time-domain analysis, Fig.~\ref{fig:error_time}, it
is made here both independent of any phase-alignment procedure and
simpler to quantify. First of all, the figure shows that the
extrapolations in radius and in resolution act in different
directions: the first one pushes the curve down (\ie less GW cycles
accumulated on a given frequency interval, tidal effects look
stronger), while the second one pushes the curve up (\ie more GW
cycles accumulated and tidal effects look weaker). This result is
qualitatively compatible with the corresponding $\Delta\phi$ curves in
Fig.~\ref{fig:error_time}, whose slopes have opposite signs. In
addition, by integrating in frequency the $\Delta Q_\omega$ curves on
the {\it common} frequency interval $MI_\omega^c =[0.045,0.067]$ one
obtains an estimate of an actual accumulated phase error that can be
compared to our previous time-domain results (\ie
Fig.~\ref{fig:error_time}). The result of this integration is given in
the seventh column of Table~\ref{tab:error_freq}. Note that the
$\Delta\phi_{\psi_4}$ computed in this way is typically significantly
larger than what was estimated above in the time domain. For instance,
regarding the comparison with the resolution extrapolated waveform,
the $Q_\omega$-based procedure indicates a phase difference of about
1.3 rad over $I_\omega^c$; by contrast, inspecting
Fig.~\ref{fig:error_time}, where the vertical (red) dashed line
corresponds to $I_\omega^c$ in the time-domain, we read from the plot
an accumulated phase difference on this interval of about 0.8 rad, \ie
about $40\%$ smaller. Similar results hold for the other phase
comparisons. This increase in the estimated phase errors is probably
due to the additional uncertainty brought by the necessity to use a
phase-cleaning procedure to compute each $Q_\omega^X(\omega)$ (see
below). This is the price we have to pay to be able to have the
convenience of an {\it intrinsic} diagnostic of the phase evolution.

A separate discussion is needed when comparing isentropic and
non-isentropic $Q_\omega$ curves. Figure~\ref{fig:error_freq}
indicates that the curve corresponding to the ideal-fluid EOS lies
above the polytropic one, and this indicates that the tidal
interaction appears {\it weaker} in the former case than the latter
(because the curve referring to the ideal-fluid is closer to the
point-mass curve than the polytropic curve, see below). This effect
was already discussed in Ref.~\cite{Baiotti08} and is likely due to
the small shocks that are formed by the interaction between the outer
layer of the stars and the external atmosphere. The polytropic EOS
should yield a priori a more accurate evolution during the inspiral,
when the stars are far apart, but should become progressively
inaccurate and inconsistent when the two stars become closer and
closer, with mass shedding and the formation of actual shocks that are
not simply due to the weak interaction with the atmosphere.
For this reason we will not use the isentropic $Q_\omega$'s as a lower
bound in our analysis, but we will focus only on non-isentropic
evolutions, though keeping in mind that there is a further source of
error on them.

A natural question that comes at this stage is: what is the
uncertainty on the determination of the $Q_\omega(\omega)$ function
that is due to the phase-cleaning (\ie phase-fitting) procedure? A
first way of addressing this issue is to measure the impact that 
changing our fiducial fitting function Eq.~\eqref{eq:phi} has on 
$Q_\omega(\omega)$. Focussing, for both models, only on our basic
${\rm IF_{HR}500}$ data, we computed the cleaned frequency using,
besides our fiducial fitting polynomial of order $n=4$ (see Eq.~\eqref{eq:phi}),
either a lower-order polynomial truncated at $n=3$ or a higher-order
one, extended up to $n=6$.\footnote{Note that $n=5$ is not meaningful
  as the corresponding $p_5$ term is exactly degenerate with
  $\phi_0$. Furthermore, the use of $x^5 \ln x$ does not help, as the
  corresponding term is nearly degenerate with $\phi_0$.} The
results of these computations are displayed in Fig.~\ref{fig:Qomg_n}
for model $\texttt{M2.9C.12}$ (top panel) and $\texttt{M3.2C.14}$
(bottom). The results are qualitatively analogous in the two
cases. First, we see that the low polynomial order $n=3$ is clearly
too small, and fails to capture (when comparing it to the PN or EOB
curves, which are accurate on the low-frequency side) the low-frequency
behavior of $Q_\omega(\omega)$. By contrast, the fact that the $n=6$
curve is barely distinguishable (on the scale of the figure) from the
$n=4$ one is an indication of a sort of ``convergence'' of our
fitting procedure as the number of $x^n$ powers is increased. We can
therefore use the {\it difference} between $Q_\omega^{n=6}(\omega)$
and $Q_\omega^{n=4}(\omega)$ as an estimate of the uncertainty $\delta
Q_\omega(\omega)$ entailed by the cleaning procedure. Computing this
difference, we find that it remains of order unity all over the
fitting frequency interval $I_\omega$. In conclusion, we estimate the
uncertainty associated with the choice of the order of the fitting 
polynomial to be $\delta Q_\omega\approx \pm 0.5$. Note that this error is rather small
compared to the various numerical errors on $Q_\omega(\omega)$
displayed in Fig.~\ref{fig:error_freq}, but it is only a lower bound
on the uncertainty level $\delta^{\rm clean} Q_\omega$ linked to the
cleaning procedure.

\begin{figure}[t]
\begin{center}
\includegraphics[width=0.45\textwidth]{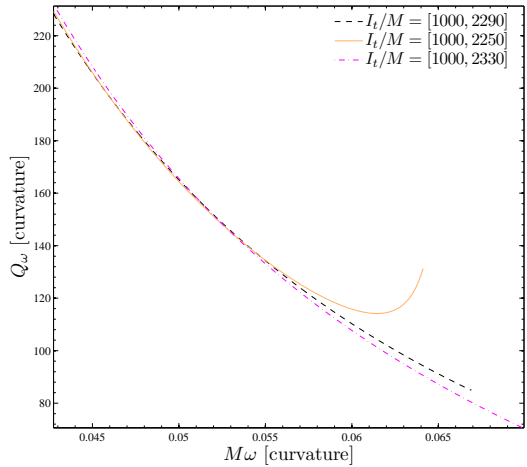}
\caption{\label{fig:Qomg_interval}Sensitivity of $Q_\omega$ to the
  choice of the fitting time-interval $I_t$ for $\texttt{M3.2C.14}$.
  Our preferred cleaning time-interval $I_t/M=[1000,\,2290]$ (central
  dashed-line) is compared to $I_t/M=[1000,\,2250]$ (solid-line) and
  $I_t/M=[1000,\,2330]$ (dash-dotted line). See text for details.}
\end{center}
\end{figure}

In particular, another relevant source of uncertainty on $Q_\omega$ is
the choice of the fitting time interval $I_t$. In
Appendix~\ref{sec:cleaning} we explicitly discuss some rules of thumb
that we follow to select $I_t$ such that the cleaning procedure is
reliable and robust. To complete the discussion of
Appendix~\ref{sec:cleaning}, we investigate (for model
$\texttt{M3.2C.14}$) the modifications in $Q_\omega$ brought by
changes in the choice of $I_t$. More precisely, we modified the
right-end point $t_2$ of our preferred cleaning time interval
$I_t/M\equiv [t_1,t_2]=[1000,\,2290]$ (see Table~\ref{tab:error_freq})
by $\pm 40$ (with fixed polynomial order $n=4$ ). The three
$Q_\omega$ curves corresponding to $t_2=\{2250,2290,2330\}$ are
displayed in Fig.~\ref{fig:Qomg_interval}. When comparing the cases
$t_2=\{2250,2290\}$, we find that the absolute value of the 
difference in $Q_\omega$ stays $\leq 1$ all over the time-interval
$I_t/M=[1000,1951]$ (corresponding to a frequency interval
$MI_\omega=[0.041,0.056]$), but then grows up to values of order $30$ near
$t_2=2250$. On the other hand, when comparing the cases
$t_2=\{2290,2330\}$ we find that the absolute value of the
difference in $Q_\omega$ stays of order $3$ all over $I_t$. This further analysis
suggests that the cleaning procedure allows us to determine $Q_\omega$
within an uncertainty level $\delta^{\rm clean} Q_\omega\approx 1$ during
most of the inspiral, with a possible increased uncertainty level
$\approx 3$ near the end of the inspiral. Note that these levels are
significantly smaller than the changes in the analytical $Q_\omega$'s
associated to a variation of the NNLO parameter $\bar{\alpha}_2$
between $0$ and $100$ (see next Section).

\section{Comparison of analytical and numerical-relativity results}
\label{sec:results}

\subsection{Characterizing tidal effects from NR simulations}
\label{sec:howto_notides}

Before proceeding with the NR/AR comparison it is useful to discuss a
procedure by means of which it is possible to effectively subtract the
tidal interaction from the NR $Q_\omega$ curves obtained so far. This
procedure will then allow us to obtain a phase diagnostic
$Q_\omega^0$ that, within some approximation, represents a non-tidally
interacting binary, namely a binary of two point-particles. As pointed
out in Ref.~\cite{Damour:2009wj}, the binding energy of a binary
system $E_b(\Omega)$ is approximately linear in $\kappa_2^T$ and it
is therefore possible to subtract the tidal effects by combining
different sets of binding-energy curves coming out  of NR
calculations. In particular, Ref.~\cite{Damour:2009wj} computed
several ``tidal-free'' binding energy curves (one curve for each
combination of two different data sets) that were compared with the
corresponding point-mass curve computed within the EOB approach or
within non-resummed PN theory. This procedure allowed for both the
identification (and thus subtraction) of systematic uncertainties in the
NR data and the discovery of higher-order tidal amplification effects.

Here we will generalize the approach introduced in
Ref.~\cite{Damour:2009wj} to the $Q_\omega$ curve. In particular we
assume that the function $Q_\omega(\omega)$ is approximately
{linear} in the (leading) tidal parameter $\k^T_2$, at least during
part of the inspiral, say up to some maximum frequency $\omega_{\rm max}$
(we will use  $\omega_{\rm max}\approx 0.07$). As a result of
this assumption, we can approximately write  $Q_\omega(\omega)$, 
for each binary, as
\begin{align}
\label{eq:Q_omega_sum}
Q_\omega(\omega;I) &= Q_\omega^0(\omega) + 
(\k_2^T)_{I}\, Q^2_\omega(\omega) + {\cal O}\left((\k_2^T)^2\right)\,,
\end{align}
where $I$ is an index labelling some binary system. As a result,
given the $Q_\omega$ diagnostics of two different binaries with labels
$(I,J)$, we can estimate the two separate functions
$Q_\omega^0(\omega)$ and $Q_\omega^2(\omega)$ as
\begin{align}
\label{eq:Qw0}
Q_\omega^0(\omega) &= \dfrac{(\k_2^T)_I Q_\omega(\omega;J) -(\k_2^T)_J
  Q_\omega(\omega;I)}{(\k_2^T)_I-(\k_2^T)_J }\,,\\
\label{eq:Qw2}
Q_\omega^2(\omega) &=
  \dfrac{Q_\omega(\omega;I)-Q_\omega(\omega;J)}{(\k_2^T)_I-(\k_2^T)_J}\, .
\end{align}
From the decomposition~\eqref{eq:Q_omega_sum}, we see that, by
definition, the function $Q_\omega^0$ denotes the $Q_\omega$
diagnostic of two {non-tidally interacting} NSs, namely of two
point-like (relativistic) masses (and also two
BHs~\cite{Damour:2009,Binnington:2009bb}). Hence, the function
$Q^2_\omega(\omega)$ is seen to represent, within the present
approximation, the effect of the tidal interaction on the $Q_\omega$
function. The calculation of both functions contains therefore
important information about the analytical representation of
tidally-interacting binary systems. In the following we will only
discuss the computation of the tidal-free part $Q_\omega^0(\omega)$,
leaving a discussion of the properties of $Q^2_\omega(\omega)$ to a
future work.

\begin{figure}[t]
\begin{center}
\includegraphics[width=0.45\textwidth]{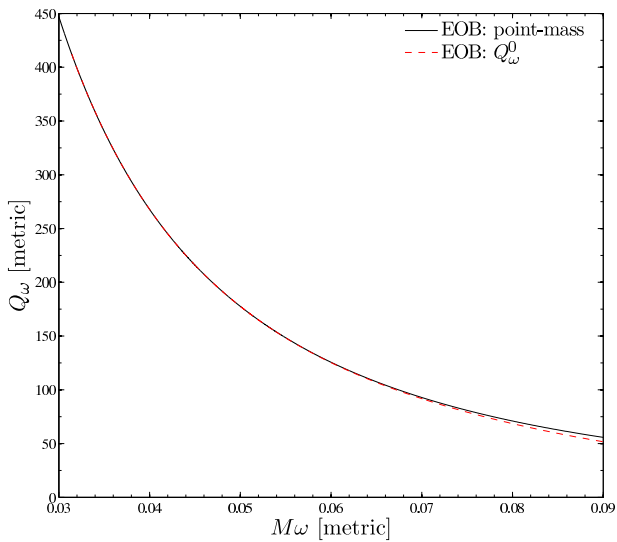}
\caption{\label{fig:Qomega0_EOB}Subtraction of tidal effects: shown as 
  a solid line is the point-mass EOB curve, while shown as a dashed line
  is the $Q_\omega^0$ curve obtained by inserting in
  Eq.~\eqref{eq:Qw0} the tidally-modified EOB $Q_\omega$ curves shown in
  Fig.~\ref{fig:explaining_Q_omega}.}
  \end{center}
\end{figure}

This subtraction procedure for computing $Q_\omega^0(\omega)$
can be first tested by using the EOB
metric waveforms computed from binaries with compactnesses ${\cal
  C}=0.12$ and ${\cal C}=0.14$. The result of the subtraction is
displayed in Fig.~\ref{fig:Qomega0_EOB}, where we compare the point-mass (\ie BBH)
EOB $Q_\omega$ curve (solid line) to the $Q_\omega^0$ curve (dashed
line), obtained by inserting in Eq.~\eqref{eq:Qw0} the ${\cal C}=0.12$ 
and ${\cal C}=0.14$ data of Fig.~\ref{fig:explaining_Q_omega}. 
The fact that the curves are barely
distinguishable up to $M\omega=0.07$ (where the difference is $\Delta Q_\omega\approx 1$)
gives us confidence that the procedure will be effective also with actual NR data. 
This will indeed be shown in the next Section.

\subsection{Inspiral: subtracting tidal effects from NR data}
\label{sec:nrQomega0}
\begin{figure}[t]
\begin{center}
\includegraphics[width=0.45\textwidth]{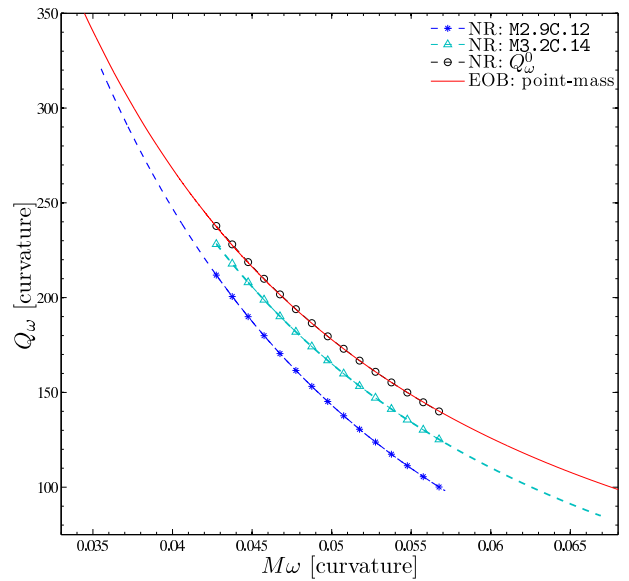}
\caption{\label{fig:Qomega0}Subtraction of tidal effects from
  numerical relativity (curvature) $Q_\omega$ curves according to
  Eq.~\eqref{eq:Qw0}. Note the excellent agreement with the
  point-mass EOB curve in the frequency window where {\tt M2.9C.12}
  and {\tt M3.2C.14} data overlap. The relative EOB-NR phase
  difference accumulated over this overlap interval is
  $\Delta\phi^{\rm EOBNR}_{\psi_4}=-0.03$~rad.}
  \end{center}
\end{figure}

We start our NR/AR comparison by computing the $Q_\omega^0$ function
as defined by Eq.~\eqref{eq:Qw0} from NR data using our two models
{\tt M2.9C.12} and {\tt M3.2C.14} as the binaries labelled $I$g and $J$ in that equation. For all the
comparisons carried out here we have limited ourselves to using the
curvature waveforms, although similar results can be obtained from the
corresponding metric waveforms.

The results are shown in Fig.~\ref{fig:Qomega0}, which reports four
different $Q_\omega$ curves: the two tidally-modified NR $Q_\omega$
curves for the binaries {\tt M2.9C.12} and {\tt M3.2C.14} (with the
asterisks and triangles highlighting a sample of the data on the
common frequency window), the subtracted NR $Q_\omega^0$ curve (with
empty circles), and the point-mass-EOB $Q_\omega$ (as a solid
line). This figure illustrates at once several of the main results of
this paper. First of all, it highlights the excellent agreement
between the cleaned NR $Q_\omega^0$ and the analytical EOB one (\cf
the red solid curve and the empty circles). This gives evidence both
for the validity of the EOB description and for the robustness of our
cleaning procedure. When we compute the relative phase difference
over the common frequency interval $[0.042,0.055]$, we obtain the
remarkably small value of $\Delta\phi^{\rm EOBNR}_{\psi_4}\equiv
\phi^{\rm EOB}-\phi^{\rm NR}=-0.03$~rad, which translates into a
relative difference $\Delta\phi^{\rm EOBNR}_{\psi_4}/\phi^{\rm
  EOBNR}_{\psi_4}=0.02\%$~\footnote{To cross-check the consistency of
  both the recovery procedure of $h_{22}$ from $\psi_4^{22}$ and
  the cleaning of the phase, we carried out the same calculation also
  for the {metric} waveforms, finding a difference $\Delta\phi^{\rm
    EOBNR}_{h}=+0.05$~rad, which is consistent with the estimated
  error-bar $\Delta\phi = \pm 0.02$~rad on the EOBNR point-mass
  waveform during inspiral~\cite{Damour:2009kr}.}. Second, it
confirms, independently of our EOB-based check (\cf
Fig.~\ref{fig:Qomega0_EOB}), that the NR tidal effects are
approximately linear in $\kappa_2^T$ at least in the early
part\footnote{In the following, we will refer to the frequency domain
  $M\omega\lesssim 0.06$ as the ``early-inspiral''. Note that for a
  fiducial $1.4\Msun-1.4\Msun$ system $M\omega=0.06$ corresponds to
  $f_{\rm GW}=690$~Hz. Note also that in the case, for instance, of
  our ${\cal C}=0.14$ system the frequency $M\omega=0.06$ is reached
  at time $t\simeq 2000M$, \ie only about $5$ GW cycles before
  merger.} of the waveform, and thus that they can be efficiently
subtracted. Third, it illustrates the fact that the tidal interaction
between the two objects is important already in the early-inspiral
part of the waveform, since both the {\tt M2.9C.12} and {\tt M3.2C.14}
curves are significantly displaced (by $\Delta Q_\omega \sim 10$) with
respect to the point-mass one. Fourth, such a good agreement with the
point-mass EOB analytical model (which was tuned so as to accurately
reproduce the equal-mass BBHs) yields an independent check of the
consistency and accuracy of our numerical simulations. Finally, we
note that in Ref.~\cite{Damour:2009wj} the procedure of subtraction,
applied there to the NR binding energy, was giving a curve slightly
displaced with respect to the point-mass EOB (or PN) curve. This
displacement was interpreted as evidence of systematic errors in the
NR simulation and prompted the introduction of a ``correcting''
procedure, which however is not necessary for the present NR data.


\begin{figure}[t]
\begin{center}
\includegraphics[width=0.45\textwidth]{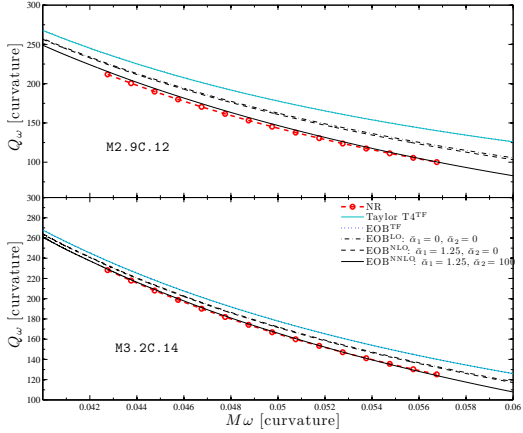}
\caption{\label{fig:Qomg_EOB}Comparison of the EOB $Q_\omega$ curves
  for different choices of the effective tidal amplification factor
  $\hat{A}_\ell^{\rm tidal}(u)=1+\bar{\alpha}_1 u + \bar{\alpha}_2
  u^2$, with the corresponding NR ones (dashed lines with open
  circles) for the two binaries considered. The dotted line
  corresponds to the ``tidal free'' (or ``point-mass'') EOB, namely,
  when ignoring tidal effects. The figure also includes the tidal-free
  Taylor-T4 model. The good visual agreement between the analytic and
  the numerical curves for $\bar{\alpha}_2=100$ provides evidence of
  the need for large NNLO tidal corrections. The corresponding phase
  differences $\Delta\phi_{\psi_4}=\phi^{\rm EOB}-\phi^{\rm NR}$ are
  listed in Table~\ref{tab:Dphi_diff}.}
  \end{center}
\end{figure}

\subsection{Early inspiral: evidence for large NNLO tidal effects}

We continue our analysis by focussing on the influence of LO tidal
effects on the early-frequency part of the $Q_\omega$ curves. We
already know from Fig.~\ref{fig:Qomega0} that tidal effects are
important in such early-frequency part of the simulations, since we
found a significant difference (of order~$10$) between the point-mass
curve and the NR ones. Can these differences be accounted just by the
LO tidal effects? Figure~\ref{fig:Qomg_EOB} shows quite clearly that
this is not the case and that the LO description {\it is not
  sufficient} to match the corresponding NR curves (dashed line with
open circles). Note that this is the case for both the {\tt M2.9C.12}
(upper panel) and the {\tt M3.2C.14} binaries (lower panel). The
difference with NR data (on the frequency interval $[0.043,0.057]$ 
where the data of {\tt M2.9C.12} and {\tt M3.2C.14} overlap)
is quantified in the first line of Table~\ref{tab:Dphi_diff} and is
rather large, namely several radians.

\begin{figure}[t]
\begin{center}
\includegraphics[width=0.45\textwidth]{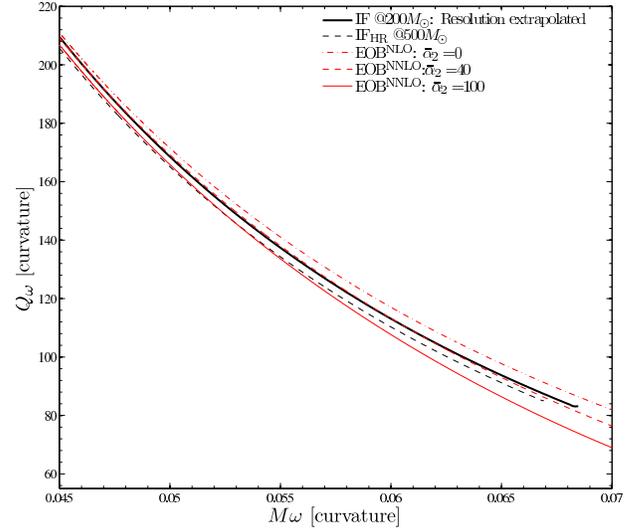}
\caption{\label{fig:span_alpha2} Magnitude of NNLO tidal effects: span
  of EOB $Q_\omega$ curves (red) with varying $\bar{\alpha}_2$ so as
  to be compatible with the various (numerical) $Q_\omega$ curves
  (black).}
  \end{center}
\end{figure}

We next turn to analyzing the effect of NLO and NNLO tidal
interactions. Here, we will regroup under the label of NLO both 1PN
and 1.5PN effects. As seen in Fig.~\ref{fig:Qomg_EOB}, the inclusion
of the NLO tidal effects ($\bar{\alpha}_1=1.25$~\cite{Damour:2009wj},
1PN tidal-radiation effects~\cite{Vines2011}, and 1.5PN tail effects)
has only a barely noticeable effect on the $Q_\omega$ curve. This
clearly indicates the need for large NNLO (2PN and higher) tidal
effects, which we chose to parameterize by means of the effective
parameter $\bar{\alpha}_2\equiv\bar{\alpha}_2^{(\ell)}$ introduced in Eq. (\ref{eq:linear2PN}). We
then found that choosing $\bar{\alpha}_2= 100$ yields a good match
between the NR and EOB $Q_\omega$ curves (solid line, EOB$^{\rm
  NNLO}$), especially for the {\tt M3.2C.14} model, for which the analytical
curve is on top of the NR data. See also Table~\ref{tab:Dphi_diff}
for the corresponding phase differences. The Table also indicates
that if we use $\bar{\alpha}_2=130$, as we did in
Ref.~\cite{Baiotti:2010}, the accumulated dephasing on the common frequency
interval $[0.043,0.057]$ is further reduced to a fraction of a
radian for both models. Note that the implementation of the EOB
waveform and radiation reaction that we use here is slightly different
with respect to the one of~\cite{Baiotti:2010}, which was based on
Ref.~\cite{Damour:2009wj} and thus did not incorporate the waveform
1PN corrections~\cite{Vines2011} nor the tail effects. This explains
why we were quoting different phase differences 
($\Delta_I\phi^{\rm EOBNR}\approx 0.1$~rad) over the same interval when 
referring to $\bar{\alpha}_2=130$ in~\cite{Baiotti:2010}. However, 
we prefer here the smaller value $\bar{\alpha}_2= 100$ because the 
corresponding $Q_\omega$ curve is, on average, closer to the NR one on 
the {\it larger frequency interval} $[0.041,0.068]$ on which we 
succeeded to clean the NR phase.

It is important to recall that various numerical errors affect the
computation of the NR $Q_\omega$ curves, and thereby affect the
quantitative determination of the effective NNLO parameter
$\bar{\alpha}_2$. For example, we have seen that the
resolution extrapolation (which seemed to be the dominant source of
uncertainty) has the practical effect of pushing the numerical
$Q_\omega$ curve {\it upwards}. This suggests that the value
$\bar{\alpha}_2 \simeq 100$ obtained from using finite-resolution NR
data is probably too large. To have a rough idea of the error range on
$\bar{\alpha}_2$, we compare in Fig.~\ref{fig:span_alpha2} various NR
and EOB curves. More precisely, this figure shows two numerical
$Q_\omega$ black curves: a solid one, derived from our fiducial
highest-resolution and largest-extraction-radius ${\rm IF_{\rm
    HR}}500$, and a dashed one, derived from the
resolution-extrapolated NR data. Also reported in
Fig.~\ref{fig:span_alpha2} are three analytical curves (red lines): namely the
EOB predictions for the three values $\bar{\alpha}_2={0,40,100}$,
respectively. Clearly, the resolution-extrapolated $Q_\omega$ curve is
close to the analytical curve corresponding to the value
$\bar{\alpha}_2 \simeq 40$, which is more than twice smaller than the
value $\bar{\alpha}_2 \simeq 100$ suggested by our fiducial,
highest-resolution NR data. It is interesting to note that the value
$\bar{\alpha}_2 \simeq 40$ agrees with the preferred value of
$\bar{\alpha}_2$ (when using $\bar{\alpha}_1=1.25$) found in
\cite{Damour:2009wj}, the work that pinpointed the first evidence for
the need of large NNLO effects. Let us also note that, independently
of the precise value of $\bar{\alpha}_2$, Fig.~\ref{fig:span_alpha2}
clearly shows the need for large NNLO effects, namely $\bar{\alpha}_2
\gtrsim 40$.

Let us also recall that the other sources of numerical error act in
various directions. For instance, non-isentropic effects actually act
so as to effectively reduce the magnitude of the tidal
interaction\footnote{Indeed the non-isentropic $Q_\omega$ curve lies
  above the isentropic one. This is
  certainly a source of error during the early-inspiral, where the
  isentropic description is a priori more accurate but some energy is
  channelled by shocks due to the interaction with the atmosphere.},
while the extrapolation to infinite extraction radius acts in the
opposite direction, namely effectively increasing the magnitude of the
tidal interaction.

In view of our incomplete knowledge of all the sources of error
intervening in our NR waveforms, we can only conclude that $\bar{\alpha}_2$
probably lies in the range $ 40 \lesssim \bar{\alpha}_2 \lesssim 130 $, with the
understanding that the lower values ($\bar{\alpha}_2 \simeq 40$) are
preferred because of the expected importance of the truncation error
in the numerical simulations. More numerical simulations with a more
detailed estimate of the numerical error budget will be needed in the
future to reduce this error range on $\bar{\alpha}_2$.

\begin{table}[t]
\caption{\label{tab:Dphi_diff}Measuring the phase difference between
  NR (curvature) waveforms and analytic ones (from both EOB and Taylor
  T4 models). The phase differences are computed on the frequency
  interval $[0.043,0.057]$ common to both $Q_\omega$ numerical
  curves. From left to right, the columns report: the type of
  analytical model, the magnitude of the effective parameters yielding
  NNLO tidal corrections; and the dephasings
  $\Delta\phi_{\psi_4}=\phi^X-\phi^{\rm NR}$ (with $X$ being either
  EOB or T4) for both {\tt M2.9C.12} and {\tt M3.2C.14} data obtained
  by direct integration of the corresponding $Q_\omega$'s of
  Figs.~\ref{fig:Qomg_EOB} and~\ref{fig:Qomg_T4} over the common 
  interval $[0.043,0.057]$.}
 \begin{center}
  \begin{ruledtabular}
  \begin{tabular}{llcc}
    Model & NNLO  & $\Delta\phi^{\tt M2.9C.12}_{\psi_4}$ & $\Delta\phi^{\tt M3.2C.14}_{\psi_4}$  \\
     &  parameters &  [rad] &  [rad]  \vspace{0.5 mm} \\
   \hline \hline
   EOB$^{\rm LO}$   & $\bar{\alpha}_2=0$    &   $5.04$   & $~1.74$ \\
   EOB$^{\rm NLO}$  & $\bar{\alpha}_2=0$    &   $4.62$   & $~1.58$ \\
   EOB$^{\rm NNLO}$ & $\bar{\alpha}_2=100$  &   $1.06$   & $~0.17$  \\
   EOB$^{\rm NNLO}$ & $\bar{\alpha}_2=130$  &   $0.056$  & $-0.25$  \\
   \hline
   T4$^{\rm LO}$    & $a_2^{\rm T4}=0$       &   $6.64$   &  $2.33$ \\
   T4$^{\rm NLO}$   & $a_2^{\rm T4}=0$        &   $6.42$   &  $2.25$ \\
   T4$^{\rm NNLO}$  & $a_2^{\rm T4}=350$      &   $1.53$   &  $0.15$ 
 \end{tabular}
\end{ruledtabular}
\end{center}
\end{table}

\begin{figure}[t]
\begin{center}
\includegraphics[width=0.45\textwidth]{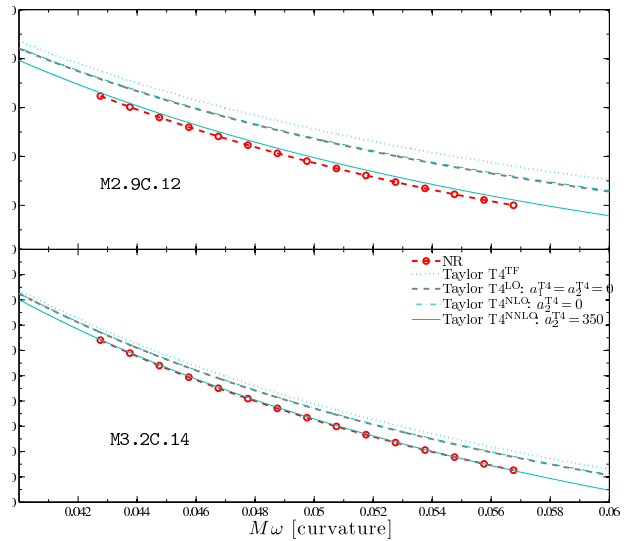}
\caption{\label{fig:Qomg_T4} Comparison of the Taylor-T4 $Q_\omega$
  curves for different choices of the effective tidal amplification
  factor $\hat{a}^{\rm tidal}(u)=1+a_1^{\rm T4} x + a_2^{\rm T4} x^2$,
  with the corresponding NR ones (dashed lines with open circles) for
  the two binaries considered. The dotted line corresponds to the
  ``tidal free'' (or ``point-mass'') T4, namely, when ignoring tidal
  effects. Note that the value $a_2^{\rm T4}=350$ of the dimensionless
  NNLO effective tidal correction parameter that best matches the
  (\texttt{M3.2C.14}) NR data is considerably larger than in the EOB
  case. The corresponding phase differences
  $\Delta\phi_{\psi_4}=\phi^{\rm T4}-\phi^{\rm NR}$ are listed in
  Table~\ref{tab:Dphi_diff}.}
  \end{center}
\end{figure}

Let us conclude this Section by briefly discussing the comparison
between the NR $Q_\omega$ diagnostics with those obtained using
several versions of the Taylor-T4 approximant. More precisely,
Fig.~\ref{fig:Qomg_T4} displays the following $Q_\omega$ curves: the
tidal-free T4 model ($T_4^{\rm TF}$, dotted line), the LO Taylor-T4
model (dashed-line), the NLO (\ie 1PN) one (dash-dotted line), and
finally the effective NNLO one (solid line), as introduced in
Sec.~\ref{sbsc:T4} above. Let us recall that the NNLO model contains
an effective 2PN parameter, called $a_2^{\rm T4}$, which is a rough T4
analog of the NNLO EOB parameter $\bar{\alpha}_2$ and which enters the
T4 tidal amplification factor Eq.~\eqref{eq:T4PN_factor}. Similarly
to the EOB case, one finds that a suitably large value of the
effective 2PN tidal parameter $a_2^{\rm T4}$ can provide
curves that are close to the numerical ones. The integrated
dephasings $\phi^{\rm T4}-\phi^{\rm EOB}$ corresponding to
Fig.~\ref{fig:Qomg_T4} are listed in Table~\ref{tab:Dphi_diff}.

\begin{figure*}[t]
\begin{center}
\includegraphics[width=0.45\textwidth]{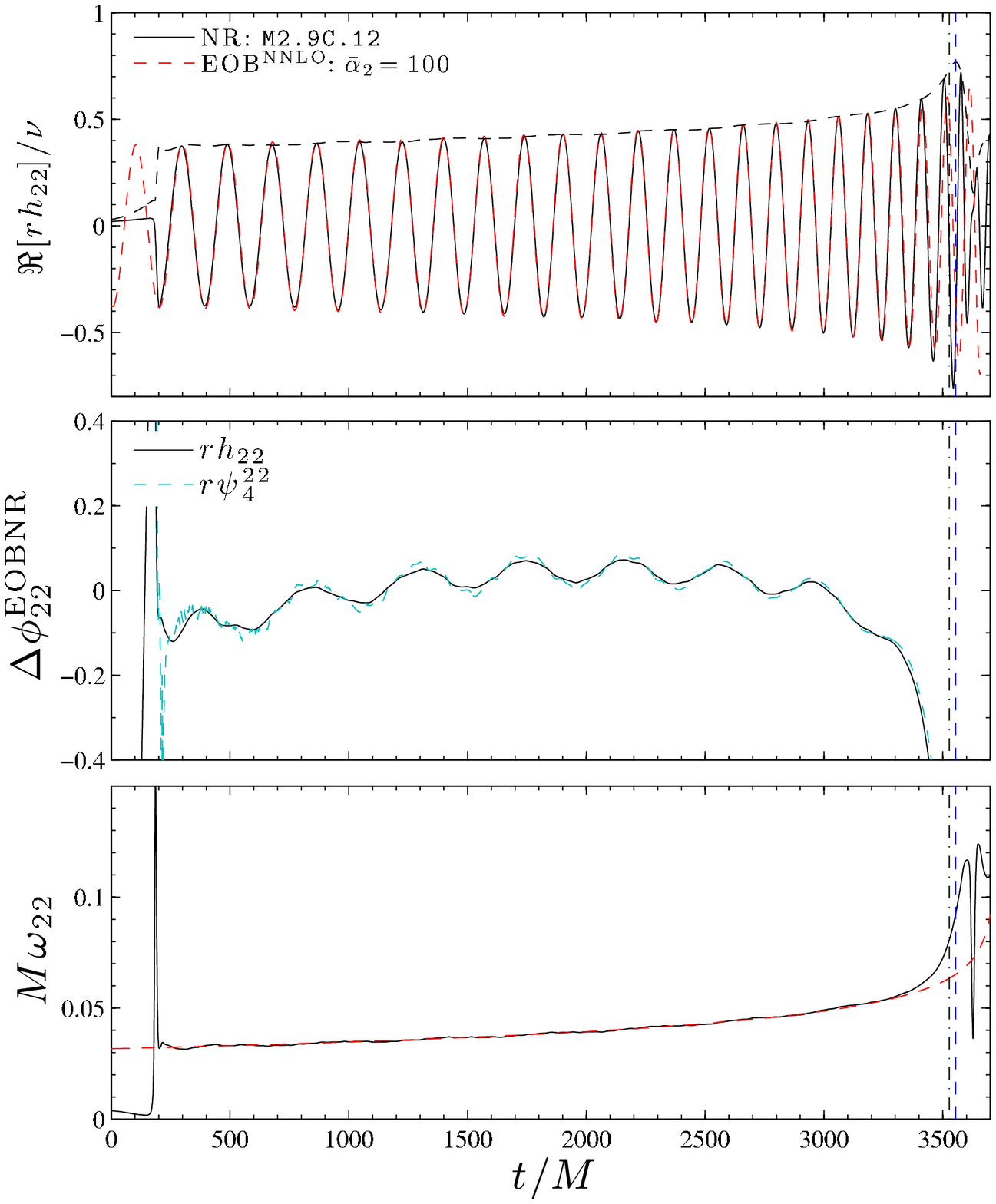}\hspace{10 mm}
\includegraphics[width=0.45\textwidth]{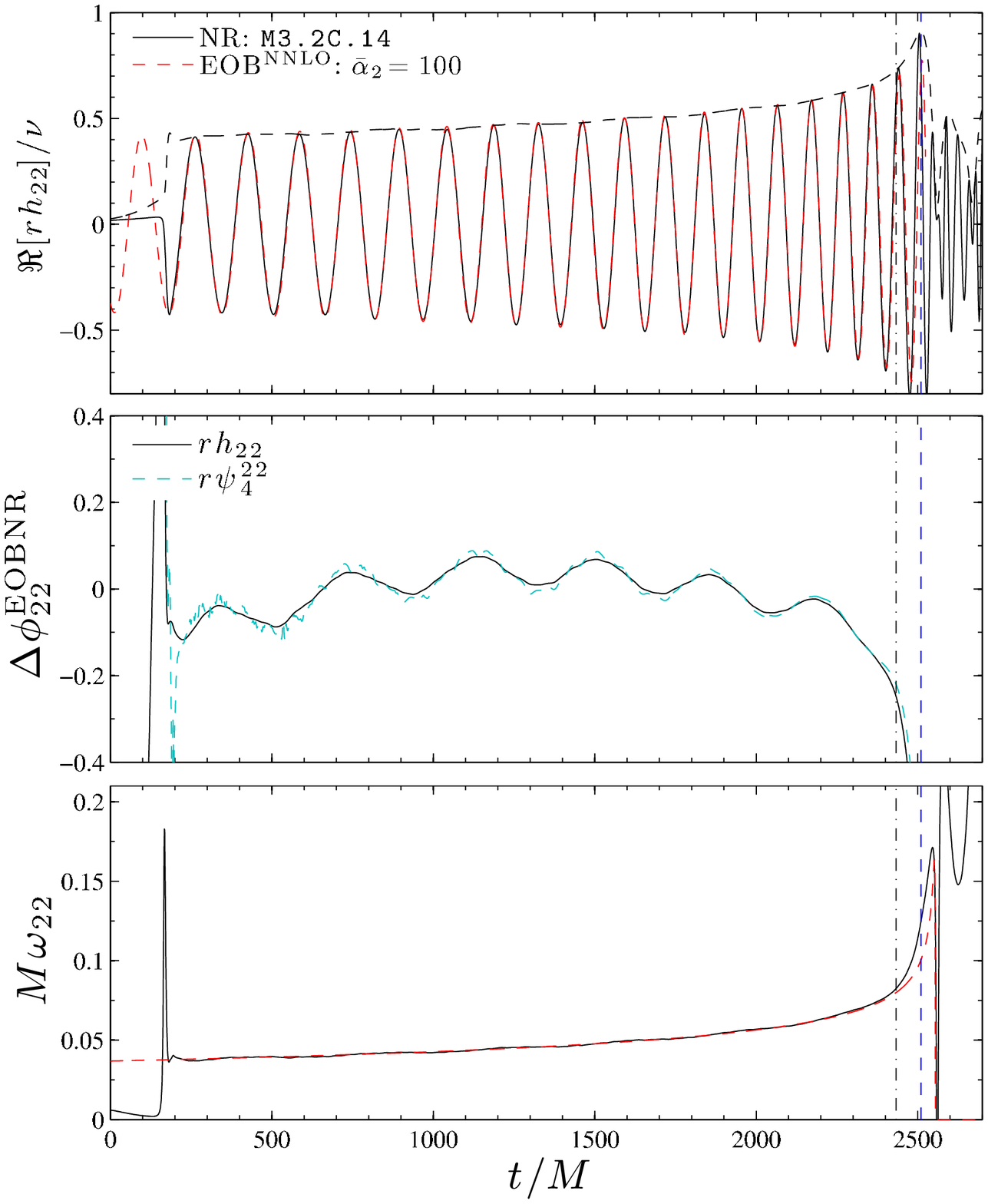}
\caption{\label{fig:eobnr_phasing_linear}Comparison between EOB and NR
  phasing for the {\tt M2.9C.12} (left panels) and {\tt M3.2C.14}
  (right panels) binaries. The top panels show the real parts of the
  EOB and NR $h_{22}$ waveforms, the middle panels display the
  corresponding phase differences $\Delta\phi^{\rm EOBNR} = \phi^{\rm
    EOB}-\phi^{\rm NR}$, both metric (solid line) and curvature
  (dashed line) waveforms, the bottom panels compare the EOB (dashed line) and
  NR (solid line) instantaneous GW frequency. The NNLO corrections to the radial potential are
  carried out with the parameter $\bar{\alpha}_2=100$. Note the
  agreement reached with the numerical waveform almost up to the time
  of the merger as defined in terms of the maximum of the GW amplitude
  (vertical dashed line) or of the contact position (dot-dashed line;
  see the text for details).}
  \end{center}
\end{figure*}

A few comments are worth making on the comparison between the EOB and
T4 results. Let us first recall that, in the BBH case, it has been
shown that the EOB description is definitely more accurate than the
Taylor-T4 one, especially when considering unequal mass
ratios~\cite{Damour:2008te} or spin effects~\cite{Hannam:2007wf}.
However, as we are considering here an equal-mass case and frequencies
that are smaller (when considering the dimensionless frequencies $ M
\omega$) than in the BBH case, the tidal-free T4 phasing is quite
close to the EOB one (see Fig.~\ref{fig:Qomg_EOB}).  Concerning
tidal-extended models, we see that both EOB and T4 approximations
highlight the need for large higher-order tidal-amplification
factors. When choosing one such amplification factor for both BNS
systems (say $\bar{\alpha}_2= 100$ for EOB and $a_2^{\rm T4}=350$ for
T4), a close look at the comparison of the corresponding $Q_\omega$
curves suggests that the EOB-predicted curves are somewhat closer than
the T4-predicted one to the NR curves. However, this, by itself, would
only be a weak indication that EOB gives a better representation of
our fiducial NR data, especially in view of the large uncertainties
discussed above on the actual value of the $Q_\omega(\omega)$
functions. On the other hand, we consider that the need of a much
larger tidal-amplification factor in the T4 case is an indication that
the analytical modelling of (LO, NLO and NNLO) tidal effects within
the EOB-resummed framework might be more robust than the corresponding
one based on Taylor-expanded approximants. Indeed, in both cases the
parametrization of NNLO effects involves multiplying tidal effects by
a factor having a similar structure: $\hat{A}^{\rm tidal
  (EOB)}_\ell(u)=1+\bar{\alpha}_1^{(\ell)} u + \bar{\alpha}_2^{(\ell)}
u^2$ versus $\hat{a}^{\rm tidal}_{\rm T4}(u)=1+a_1^{\rm T4} x +
a_2^{\rm T4} x^2$. In addition, the quantities $u$ and $x$ are
numerically close to each other (both being close to $(M
\omega/2)^{2/3} \sim v^2/c^2$). At the end of the inspiral, $M \omega$
reaches numerical values of order $0.1$ (\ie $1154$ Hz for a fiducial
BNS system), corresponding to $u \simeq x \simeq 0.136$. For such a
value one sees that the EOB amplification factor (with
$\bar{\alpha}_2= 100$) remains relatively moderate\footnote{For
  $\bar{\alpha}_2= 40$, this amplification factor becomes
  $\hat{A}^{\rm tidal (EOB)}_\ell(u)= 1 + 1.25 u + 40 u^2 \simeq 1+
  0.17 + 0.74 \simeq 1.91$}, namely $\hat{A}^{\rm tidal
  (EOB)}_\ell(u)= 1 + 1.25 u + 100 u^2 \simeq 1+ 0.17 + 1.85 \simeq
3$, while the T4 one (with $a_2^{\rm T4}=350$) is suspiciously large,
and is completely dominated by the 2PN contribution, namely
$\hat{a}^{\rm tidal}_{\rm T4}(u)=1+ 1.19 x + 350 x^2 = 1 + 0.16 + 6.47
= 7.63$. Another way to phrase this is to notice that the large T4
value $a_2^{\rm T4}=350$ is such that the 2PN contribution $a_2^{\rm
  T4} x^2$ starts dominating the LO term at $x=1/\sqrt{350} \simeq
1/18.7$, \ie at large separations $ r \simeq 18.7 M$ corresponding to
rather low frequencies $M \omega = 2 x^{3/2} = 0.025$, \ie $285$ Hz
for a fiducial BNS system. Furthermore, such a large value for
$a_2^{\rm T4}$ works well for binary {\tt M3.2C.14}, but less well for
binary {\tt M2.9C.12}.

Clearly, in view of the large current uncertainties on the $Q_\omega$
NR curve, more work is needed to confirm this provisional
conclusion. In particular, more accurate NR simulations, encompassing
more compactnesses and different mass ratios will be needed to assess
the relative merits of the EOB versus the Taylor-T4 description of
tidally interacting BNS systems.

\subsection{EOB/NR phasing}
\label{sec:eobnr_phasing}

So far our NR/AR comparison based on the function $Q_\omega(\omega)$
has been limited to a frequency interval which did not cover the last
octave of frequency evolution, even if, when viewed in the time
domain, this interval covered most of the cycles of the inspiral. In
this section we finally focus on a phasing comparison in the time
domain which covers {\it the full inspiral and plunge phase}, up to
the merger of the two NSs. Our strategy here will not be to explore
from scratch a good range of values of the tidal NNLO parameter
$\bar{\alpha}_2$ values, but instead to use the value
$\bar{\alpha}_2=100$ suggested by our previous
$Q_\omega(\omega)$-analysis, and to explore to what extent it succeeds
in providing a waveform which agrees with our fiducial
(highest-resolution) NR waveform over the full inspiral. Anticipating
our conclusion, we will find that the EOB waveform with
$\bar{\alpha}_2=100$ does closely agree (both in phase and modulus)
with the NR waveform essentially up to the merger.

This is shown in Fig.~\ref{fig:eobnr_phasing_linear}, which compares
the (real part of the) EOB and NR metric $rh_{22}$ waveforms for the
case including NNLO effects with $\bar{\alpha}_2=100$. The left panels
refer to the {\tt M2.9C.12} binary, while the right panels refer to
the {\tt M3.2C.14} one. The top panels show the real parts of both the
EOB and NR $h_{22}$ waveforms (divided by the symmetric mass ratio
$\nu$); the middle panels display instead the corresponding phase
differences $\Delta\phi^{\rm EOBNR}(t) = \phi^{\rm EOB}(t)-\phi^{\rm
  NR}(t)$, for both metric (solid line) and curvature (dashed line) waveforms,
for completeness; the bottom panels compare the EOB (dashed line) and
NR (solid line) instantaneous GW frequency. The least-squares phase
alignment has been performed on the time interval $[t_L,t_R]/M = [250,
  3300]$ for the {\tt M2.9C.12} binary and $[t_L,t_R]/M = [250, 2250]$
for the {\tt M3.2C.14} one.

The two vertical lines (dot-dashed and dashed) indicate the ``end of
the inspiral phase'', as defined either within the EOB analytical
framework (dot-dashed line) or by using NR information (dashed line).
Note that we call here simply ``inspiral'' what was called
``insplunge'' in previous EOB studies, namely the union of the
inspiral and (when it is reached before merger) of the plunge. More
precisely, the dashed line indicates the NR-defined ``merger'', \ie
the time (computed from the NR data) at which the modulus of the
metric waveform reaches its first maximum. On the other hand, the
vertical dash-dotted line indicates the EOB-defined ``contact''
between the two NSs\footnote{Note that the styles of the corresponding
merger and contact vertical lines as depicted in the two panels 
of Fig.~2 of Ref.~\cite{Baiotti:2010}
are inverted with respect to the text there. 
See the arXiv version for the correct figures.}. 
Such a formal contact moment was introduced in
Eqs.~(72) and (77) of Ref.~\cite{Damour:2009wj}, by a condition
expressing that the EOB radial separation $R$ becomes equal to the sum
of the tidally deformed radii of the two NSs, namely
\begin{equation}
R^{\rm contact} = (1+h_2^A \, \epsilon_A (R^{\rm contact})) \, R_A + \left\{
~_A~\leftrightarrow~_B\right\} \, , 
\end{equation}
where $\epsilon_A(R) = {M_B}{R_A^3}/({R^3}{M_A})$ is the dimensionless
parameter controlling the (LO) strength of the tidal deformation of
the NS labeled $A$ by its companion $B$ and where $h_2^{A,B}$ is the
shape Love number~\cite{Damour:2009,Damour:2009va}. A recent study of
the tidally induced shape deformation of BHs~\cite{Damour:2009va} has
shown that the BH shape Love number $h_2$ was a function of the
separation $r$, which increased as $r$ decreased (and $u$
increased). This behavior is similar to the behavior of the
(effective) quadrupole Love number $k_2^{\rm eff}(u) = k_2(
1+\alpha_1^{(2)} u + \alpha_2^{(2)} u^2)$, where both
$\alpha_1^{(2)}$ and $\alpha_2^{(2)}$ were found to be
positive~\cite{Damour:2009wj}.  One would need a special study devoted
to the comparison of the EOB-predicted NS shape deformation to NR data
to investigate in detail the $u$ dependence of the analogous $h_2^{\rm
  eff}(u) = h_2( 1+\gamma_1^{(2)} u + \gamma_2^{(2)}
u^2)$. Leaving to future work such a study, we will here replace the
$u$-dependent effective shape Love number $h_2^{\rm eff}(u)$ by a
constant, chosen such that the EOB-predicted contact happens {\it
  before} the NR-defined merger for the two BNS systems we
consider. We found that $h_2^{\rm eff}=3$ works, and this is the value
we will use to replace $h_2^A$ and $h_2^B$ in the contact condition
written above\footnote{A similar approach was taken
  in~\cite{Damour:2009,Baiotti:2010}, with a less conservative value 
$h_2^{\rm eff}=1$. Let us recall that the computation of the
  infinite-separation shape Love number $h_2=h_2^{\rm eff}(u=0)$ of
  NSs gives values of order unity~\cite{Damour:2009}.}. An important
point to note is that our (EOB-based) analytical definition of contact
allows one to analytically predict a complete inspiral waveform,
including its termination just before merger.

Figure~\ref{fig:eobnr_phasing_linear} shows that the agreement in the
time domain between the analytic EOB description and the fully
numerical one is extremely good essentially up to the merger.  More
precisely, the match between the two descriptions is excellent both in
modulus and in phase, with a dephasing of order $\Delta\phi= \pm 0.1$
rad during most of the long inspiral phase. It is only during the last
$100M$ before contact that the dephasing grows significantly.  One
should note that this excellent EOB/NR agreement holds for
\textit{both} binaries {\tt M3.2C.14} and {\tt M2.9C.12}, and has been
obtained by tuning a {\it single} tidal-amplification parameter.

Clearly the results presented here give only a first cut at these issues.
More NR/AR comparisons are needed to confirm our findings and to
determine the most effective value of $\bar{\alpha}_2$. With
sufficiently accurate NR data one can hope to determine not only
the effective tidal-amplification factor
$\hat A^{\rm tidal}_\ell(u) = 1+\bar{\alpha}_1 u + \bar{\alpha}_2 u^2$,
but also the precise separation-dependence of $\hat A_\ell^{\rm tidal}(u)$. 
This would allow one to extend the EOB description right up to the merger.

\section{Conclusions}
\label{sec:end}

We have presented the first comprehensive NR/AR comparison of the
gravitational waveforms emitted during the inspiral of relativistic
BNSs as computed via state-of-the-art
numerical-relativity simulations and as modelled via state-of-the-art
analytical approaches. Overall, the work reported here and our
findings can be summarized as follows.

\begin{enumerate}

\item We have considered the longest to date numerical simulations of
  inspiralling and coalescing equal-mass BNSs modeled either with an
  ideal-fluid or a polytropic EOS. Because tidal effects are most
  sensitive to the stellar compactness, we have considered two
  binaries with either a small compactness of ${\cal C}=0.1199$ or
  with a large compactness of ${\cal C}=0.1396$. The parts of the
  waveforms relative to the inspiral cover between $20$ and $22$
  cycles and have been studied to isolate possible sources of error,
  such as non-isentropic evolutions, finite-radii GW extractions, and
  the use of finite resolutions. For the model with the highest
  compactness, the first two sources of errors lead to a total
  error-bar in the GW phase of $\Delta\phi\simeq \pm 0.15$~rad, while
  the error coming from a finite resolution indicates an accumulated
  phase error of $\Delta\phi\simeq \pm 0.54$~rad.

\item We have used the function $Q_\omega(\omega)\equiv
  \omega^2/\dot{\omega}$ as a useful diagnostic of the physics driving
  the evolution of the GW frequency $\omega$. The calculation of this
  quantity is however challenging when made from the early-inspiral
  part of the NR waveforms, as the latter is affected by a series of
  contaminating errors. We have filtered out these errors by fitting
  the NR phase evolution $\phi(t)$ with a simple analytical expression
  that reproduces at lower order the behavior expected from the PN
  approximation. We have compared the various $Q_\omega$'s obtained
  from different data to estimate the error range entailed by
  comparing analytical predictions to our highest-resolution,
  largest-extraction-radius NR data.

\item Using the estimated $Q_\omega(\omega)$ function we have shown
  that it is possible, at least for frequencies $M\omega\lesssim 0.06$
  (\ie $f_{\rm GW}\lesssim 700$~Hz for a fiducial $1.4\Msun-1.4\Msun$
  BNS system), to {\it subtract the tidal-effect contribution} from
  the NR waveforms and consistently match this with the expected EOB
  model for point particles which has been successfully matched to BBH
  simulations. The ability to perform this match accurately provides
  us with an independent validation of the quality of our numerical
  results as well as with a confirmation that the function
  $Q_\omega(\omega)$ is approximately linear in the (leading) tidal
  parameter $\k^T_2$.

\item The comparison of analytical predictions with NR data shows that
  tidal effects are significantly amplified by higher-order (NNLO)
  relativistic corrections even in the early inspiral phase. These
  NNLO tidal corrections are parameterized within the EOB approach by a
  unique (effective, 2PN) tidal parameter $\bar{\alpha}_2$. Although
  the most precise available at the moment, the quality of the NR data
  is such that we can only constrain the actual value of
  $\bar{\alpha}_2$ to be in the range $ 40 \lesssim \bar{\alpha}_2
  \lesssim 130$.

\item Once a \textit{single} choice for $\bar{\alpha}_2$ is made, the
  EOB-predicted waveforms agree (both in phase and in modulus) with
  the NR ones (for both BNS systems) within their error bar and
  essentially up to the merger.

\item Finally, we have also compared the NR phasing with the one
  predicted by a non-resummed Taylor-T4 PN expansion, completed by
  additional tidal terms. If one uses only the currently known
  analytic T4 tidal terms, the T4 model dephases (when $\C=0.12$) by
  more than $2 \pi$~rad already at the GW frequency $ M \omega =
  0.057$, which is about twice smaller than the GW frequency at merger
  (we recall that $M \omega = 0.057$ corresponds to $658$ Hz for a
  fiducial $1.4 \Msun - 1.4 \Msun$ system). On the other hand, a good
  match (for both compactnesses) with the NR phasing is possible if
  one allows for a T4 analog of the EOB $\bar{\alpha}_2$ parameter,
  \ie an (effective) 2PN amplification of tidal effects. The
  corresponding parameter $a_2^{\rm T4} \simeq 350$ is suspiciously
  large, works well for binary {\tt M3.2C.14} but less well for binary
  {\tt M2.9C.12}, and dominates the amplification of tidal effects
  already at frequencies $ M \omega = 0.025$ (corresponding to $285$
  Hz). This seems to suggest that the EOB-based representation of
  tidal effects is more reliable than the Taylor-T4 one.

\end{enumerate}

In summary, the work presented here opens new avenues to the important
synergy between numerical and analytic descriptions of inspiralling
compact-object binaries in general relativity. For the first time we
have shown that an analytic modelling is possible also for objects
which cannot be treated as point-particles and for which, therefore,
tidal effects represent important corrections. Although the results
presented here are very encouraging, a number of improvements are
needed on both the numerical and the analytical sides. On the
numerical side, higher resolutions and better measures of the
convergence rates (which are particularly challenging in non-vacuum
simulations) are needed to decrease the numerical phase errors to and
reach firm conclusions about the tidal contributions to the phasing.
On the analytical side, higher-order PN calculations are needed to
better determine the form of the NNLO corrections. Both of these goals
will be the subject of our future work. Hopefully, progress on both
fronts will enable us to determine the crucial tidal-induced dephasing
function $\Delta^{\rm tidal}\phi(\omega)$ with an accuracy
sufficiently high to extract reliable information on the EOS of matter
at nuclear densities~\footnote{Simple estimates based on the scaling
  $\kappa_2^T\propto R^5$ suggest that one needs to know $\Delta^{\rm
    tidal}\phi(\omega)$ with a fractional accuracy better than $20\%$
  to constrain NS radii to a relative precision of $\delta R/R\approx
  4\%$. See also Ref.~\cite{Pannarale2011} for the prospects from a
  BH-NS system.}.

\begin{acknowledgments}

We are grateful to Sebastiano Bernuzzi for discussion throughout the
development of this work and to Francesco Pannarale for useful
input. The simulations were performed on the Ranger cluster at the
Texas Advanced Computing Center through TERAGRID allocation
TG-MCA02N014, and on the cluster Damiana of the AEI. This work was
supported in part by the DFG Grant SFB/Transregio 7, by ``CompStar'',
a Research Networking Programme of the European Science Foundation, by
the JSPS Grant-in-Aid for Scientific Research (19-07803), by the MEXT
Grant-in-Aid for Young Scientists (22740163), and by NASA Grant
No. NNX09AI75G.

\end{acknowledgments}

\appendix

\section{Computing metric waveforms from $\psi_4$}
\label{appendix_A}

We  discuss here the details of how to accurately derive the
metric waveforms $h_{+,\times}$ from the numerically computed
curvature waveforms $\psi_4$. We first recall that the procedure
outlined in Ref.~\cite{Damour:2008te} consisted essentially of three
steps. (i) First one performs the double integration of $\psi_4^{\ell m}$
starting at $t=0$ with the integration constants set to zero; this amounts to defining
\begin{align}
\label{eq:hdot}
\dot{h}_0^{\ell m}(t) &\equiv \int_0^t dt' \psi_4^{\ell m}(t'),\\
\label{eq:hraw}
h_0^{\ell m}(t)       &\equiv \int_0^t dt'\dot{h}_0^{\ell m}(t').
\end{align}
The provisional metric waveform $h_0^{\lm}$ differs from the ``exact''
metric waveform \eqref{eq:int_h_infty} (integrated from past infinity)
by a linear function of $t$, say
\begin{equation}
\label{eq:h0_exact}
h_0^{\lm}(t) = h^{\lm}(t) + \alpha_{\rm exact}t + \beta_{\rm exact}.
\end{equation}
(ii) The second step consists in obtaining an estimate of the two
(complex) integration constants ($\alpha_{\rm exact},\beta_{\rm
  exact}$) that enter the metric waveform~\eqref{eq:int_h_infty} by
fitting over the {full} simulation time interval the $(t\geq 0)$-integrated
waveform~\eqref{eq:hraw} to a linear function of $t$, say $h_0^{\rm
  lin-fit}=\alpha t + \beta$, where $\alpha$ and $\beta$ are complex
quantities. (iii) The third and final step of the procedure of
Ref.~\cite{Damour:2008te} consisted in subtracting the linear function
$\alpha t+\beta$ from $h_0^{\ell m}$ so as to define an approximation
to the ($t\geq -\infty$)-integrated metric waveform, say $h_{\ell
  m}^{\rm old}(t) \equiv h_0^{\ell m}(t)-h_0^{\rm lin-fit}(t)$.

Here we will use a ``new'' (three-step) procedure, which starts with
the same step (i), but modifies both steps (ii) and (iii) so as to get
a better approximation to the exact metric waveform. First of all, we
define an ``adiabatic-like'' approximation to the metric waveform,
\begin{equation}
\tilde{h}_{\lm}(t)\equiv-\dfrac{\psi_4^{\ell m}(t)}{\omega_{\lm}^2(t)}
\end{equation}
and use this to define
\be
\tilde{h}^{\lm}_0(t) \equiv h_0^{\lm}(t)-\tilde{h}_{\lm}(t).
\end{equation}
As $\tilde{h}^{\lm}(t)$ is approximately equal to $h^{\lm}(t)$
(because of the approximately adiabatic nature of the inspiral), we
see from Eq.~\eqref{eq:h0_exact} that
$\tilde{h}_0^{\lm}(t)=h_{\lm}(t)-\tilde{h}_{\lm}(t)+\alpha_{\rm exact}
t + \beta_{\rm exact}$ will be closer to the unknown linear function
$\alpha_{\rm exact}t + \beta_{\rm exact}$ than $h_0^{\lm}(t)$ was.
Therefore, the next step is to perform the linear fit on this
$\tilde{h}^{\lm}_0$ rather than on $h_0^{\lm}(t)$ itself. Then, the
last step (iii) consists, as in the past, in subtracting the resulting
improved linear fit $\alpha t + \beta$ from the ($t\geq
0)$--integrated metric waveform $h_0^{\ell m}(t)$.

In addition, let us note that we perform the fit not on the whole time
interval, but rather on a restricted time interval that cuts away the
first cycles of the waveform. Finally, after doing several tests, we
realized that the entire procedure leads to a physically more reliable
metric waveform (see below) if $\tilde{h}^{\lm}_0(t)$ is fitted not to
a simple {linear} function, but rather to
a {quadratic}\footnote{We think that such a quadratic fit is needed for 
absorbing several effects that ``pollute'' the waveform,
notably finite-extraction-radius effects, remnant junk radiation, etc. 
In this respect, we also mention that Ref.~\cite{Baiotti:2008nf}, 
in the context of non-spherical star oscillations, found that a 
quadratic polynomial used in the recovery of $h_{20}$ from $\psi_4^{20}$ 
was a necessary choice to find a good agreement both with
Abrahams-Price metric extraction and perturbative waveforms.}  
one, $h_0^{\rm quad-fit}(t)= \gamma t^2 + \alpha t + \beta$.

As emphasized in Ref.~\cite{Damour:2008te}, we accept the integrated
waveform {if and only if} its modulus exhibits a monotonic growth in
time during the inspiral, consistently with the expected circularly
polarized behavior of the metric waveform (as well as the curvature
one)\footnote{Note however that small-amplitude, high frequency
  ``ripples'' {are} still present in the modulus. Their origin is
  however essentially numerical, as they are {also} present in the
  modulus of $\psi_4^{22}$.}. Figure~\ref{fig:fig_metric_waves}
displays the metric waveforms (both for the {\tt M2.9C.12} (left) and
the {\tt M3.2C.14} (right) models) obtained using this improved
procedure. The time intervals where we fit the waveforms to get
$h_0^{\rm quad-fit}(t)$ start respectively at $t_1/M=294$ (model {\tt
  M2.9C.12}) and at $t_1/M=677$ (model {\tt M3.2C.14}). Note how the
modulus of both models exhibits a smooth monotonic behavior in time.

\section{Cleaning the GW phase and $Q_\omega$ curves}
\label{sec:cleaning}

\begin{figure*}[t]
\begin{center}
\includegraphics[width=0.9\textwidth]{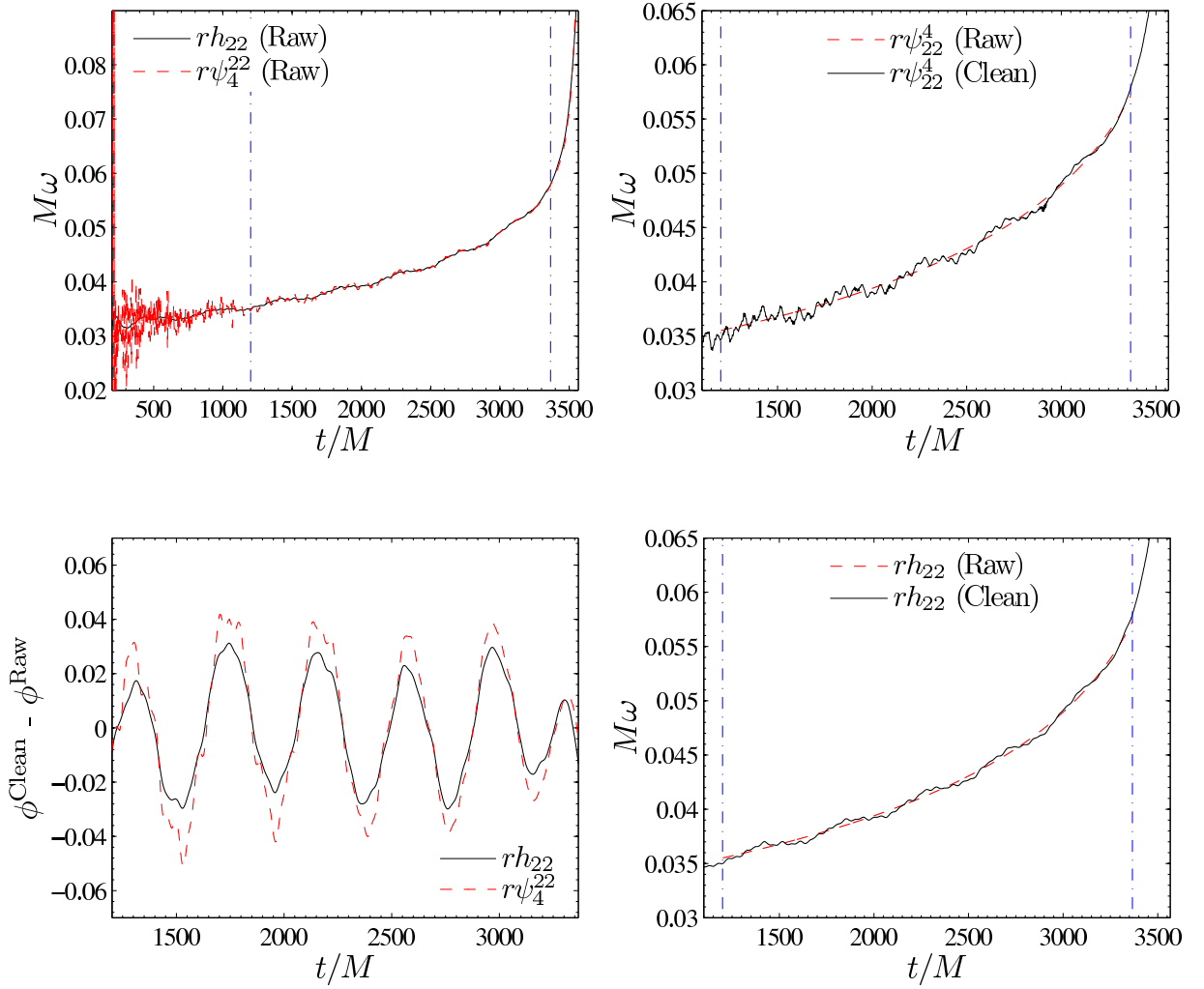}
\caption{\label{fig:test_M2.9C.12}Testing the fit of the GW phase of
  the {\tt M2.9C.12} simulation. The top-left
  panel shows the time evolution of the frequency, computed from the metric and curvature
  waveforms. The bottom-left panel shows the deviation of the cleaned phase evolution with respect to
  the raw data; note that they average to zero. The right panels show the comparison of the frequency
  evolution of the cleaned and raw waveforms, for the curvature (top) and metric (bottom) waveforms.}
  \end{center}
\end{figure*}

\begin{figure*}[t]
\begin{center}
\includegraphics[width=0.9\textwidth]{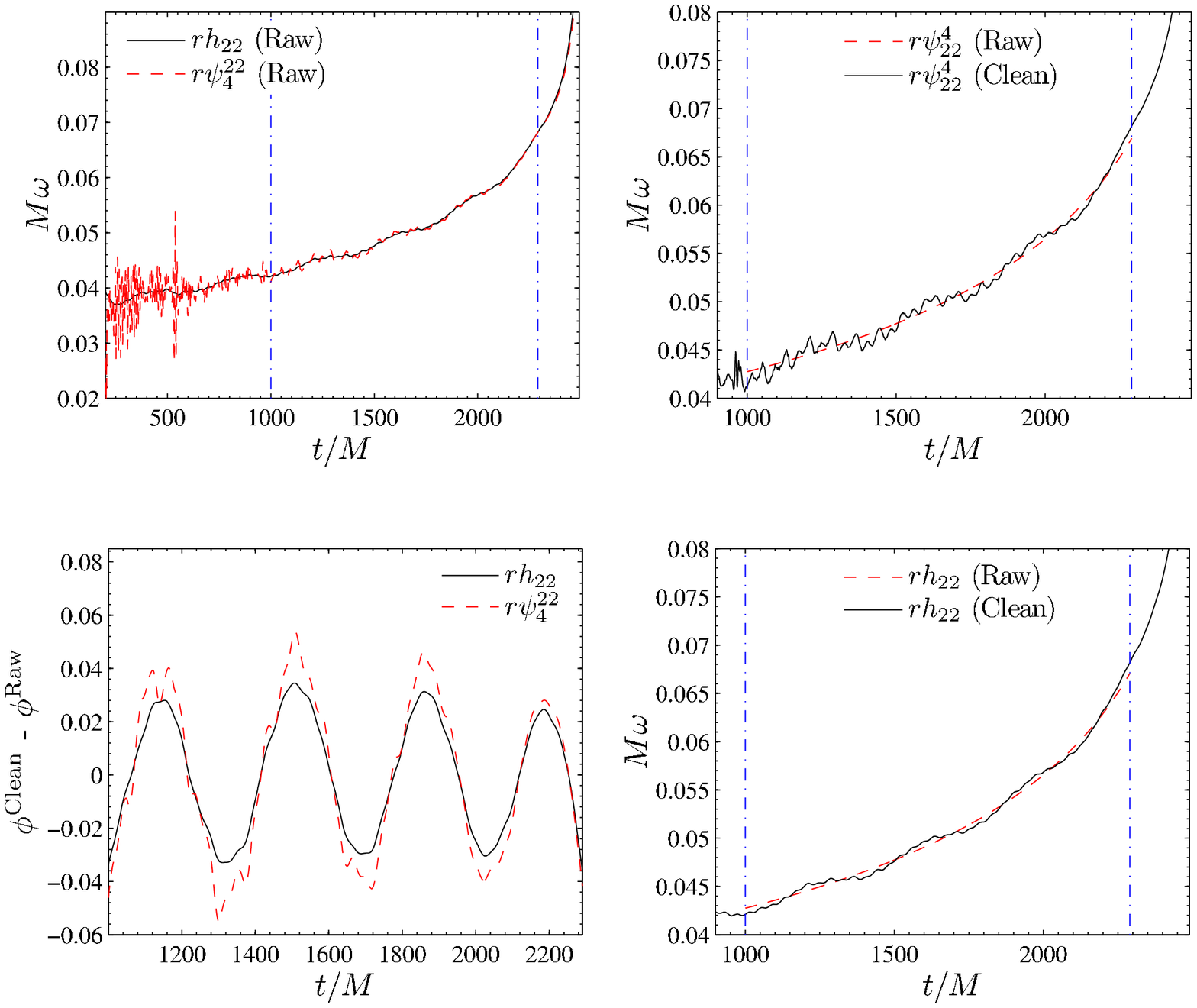}
\caption{\label{fig:test_M3.2C.14}The same as Fig.~\ref{fig:test_M2.9C.12} but for the {\tt M3.2C.14} simulation.}
  \end{center}
\end{figure*}

We next provide more detailed information about the cleaning procedure
of the NR GW phase advocated in Sec.~\ref{sec:Qomega} and used to
drive NR/AR comparisons. As we said in the main text, the final goal
is to fit away the high-frequencies oscillations in the GW phase
$\phi$ so as to get a clean and smooth $Q_\omega$ curve,
Eq.~\eqref{eq:5.15}. We recall that the idea is to fit $\phi(t)$ with
an analytic expression that is modeled on the PN expansion. Defining
the quantity
 
\begin{equation}
x(t, \phi_c)=\left\{\dfrac{\nu}{5}(t_c -t)\right\}^{-1/8}, 
\end{equation} 
one then fits the NR phase with an expression of the form
\begin{equation}
\label{eq:fit_phi}
\phi = -\dfrac{2}{\nu}x^{-5}\left(1 + p_2 x^2 + p_3 x^3 + p_4 x^4 + \dots \right)+\phi_0,
\end{equation}
where $t_c$, $\phi_0$, and the $p_i$'s are free coefficients to be
determined by the fit. Note that $t_c$ can be thought of as defining
a formal ``coalescence'' time. There are two delicate (correlated)
points: (i) how many powers of $x$ [possibly including also
  $x^n\ln(x)$ terms] one has to include in Eq.~\eqref{eq:fit_phi}, and
(ii) on which (time) interval $I_t/M=(t_1,t_2)$ the approximate
description of $\phi$ given by Eq.~\eqref{eq:fit_phi} (and
consequently of $Q_\omega$) is reliable. The procedure to select the
``best'' time interval and to consistently assess the quality of our
cleaned curves can be summarized as follows:

\begin{enumerate}

\item The initial time $t_1$ is chosen so as to eliminate as much as
  possible the most noisy part of the curvature frequency. In practical
  terms, this meant cutting at $t_1=1200$ for {\tt M2.9C.12} data and
  $t_1=1000$ for ${\tt M3.2C.14}$ data. This is illustrated in the
  top-left panels of Fig.~\ref{fig:test_M2.9C.12} (for {\tt M2.9C.12}
  data) and of Fig.~\ref{fig:test_M3.2C.14} (for {\tt M3.2C.14} data),
  which show the curvature (dashed line) and metric (solid line)
  instantaneous GW frequency $\omega$. In both plots, the first
  vertical line identifies the location of $t_1$.

\item For a given order of the polynomial, we found the right end,
  $t_2$, of the time window essentially, by trial and error,
  monitoring the behavior of several quantities. In particular, (i) we
  checked that the cleaned $\omega$ visually ``averages'' the raw
  $\omega$, for both $\psi_4^{22}$ and $h_{22}$ data. This is
  illustrated in the top-right and bottom-right panels of
  Figs.~\ref{fig:test_M2.9C.12}-\ref{fig:test_M3.2C.14}, the raw data
  appearing as dashed lines, the cleaned data as solid lines. Then,
  (ii), we require that the phase difference $\phi^{\rm
    Clean}-\phi^{\rm Raw}$ averages to zero, which indicates that we
  have subtracted all the ``secular'' trends by means of our
  polynomial fit. The quantity $\Delta\phi^{\rm CleanRaw}=\phi^{\rm
    Clean}-\phi^{\rm Raw}$ (both curvature and metric) is displayed in
  the bottom-left panel of
  Figs.~\ref{fig:test_M2.9C.12}-\ref{fig:test_M3.2C.14}. The fact that
  it averages to zero is the indication that our fit caught the
  ``secular'' behavior of the phase, averaging away both (numerical)
  low-frequency and high-frequency oscillations.

\item For a fixed time window, the inspection of $\Delta\phi^{\rm
  CleanRaw}$ is also crucial for choosing the order of the polynomial
  in $x$, which we set to be of fourth-order. A 3rd-order one is
  clearly not enough to get the right trend of the frequency (and thus
  of $Q_\omega$) up to the end of our preferred interval.

\item To better select the end $t_2$ of the time window, we found it
  useful to monitor the difference between the curvature and metric
  $Q_\omega$'s, namely $\Delta Q_\omega^{\rm c-m}=Q_\omega^{\rm
    curvature}-Q_\omega^{\rm metric}$. We typically choose the value
  of $t_R$ in such a way that $\Delta Q_\omega^{\rm c-m}$ is always
  smaller than $0.2$ on the frequency interval corresponding to
  $I_t/M$. This value can be estimated by comparing curvature and
  metric $Q_\omega$'s within the EOB. For example, for the NNLO model
  with $\bar{\alpha}_2=100$ one checks that $\Delta Q_\omega^{\rm
    c-m}\lesssim 0.2$ when $\omega\in[0.035,0.055]$ for ${\cal
    C}=0.12$, and $\Delta Q_\omega^{\rm c-m}\lesssim 0.2$ when
  $\omega\in[0.035,0.063]$ for ${\cal C}=0.14$. This gives us an idea
  of the level of $\Delta Q_\omega^{\rm c-m}$ that we can accept from
  our cleaned NR curves, so that we can choose the fitting time window
  accordingly.

\end{enumerate}

In conclusion, to obtain the central NR-cleaned $Q_\omega$ curves
labelled ${\rm IF_{HR}}500$ used in the core of the paper, we fixed
$t_R/M=3366$ for the {\tt M2.9C.12} phase and $t_R/M=2290$ for the
{\tt M3.2C.14} one. The time intervals (and the corresponding
frequency ones) used to clean the other NR phases are also listed in
Table~\ref{tab:error_freq}.

\bibliography{AEIBibtex/aeireferences}

\end{document}